# Dark matter vorticity and velocity dispersion from truncated Dyson–Schwinger equations

**Alaric Erschfeld and Stefan Floerchinger**

Theoretisch-Physikalisches Institut, Friedrich-Schiller-Universität Jena
Fröbelstieg 1, D-07743 Jena, Germany

E-mail: alaric.erschfeld@uni-jena.de, stefan.floerchinger@uni-jena.de

**Abstract.** Large-scale structure formation is studied in a kinetic theory approach, extending the standard perfect pressureless fluid description for dark matter by including the velocity dispersion tensor as a dynamical degree of freedom. The evolution of power spectra for density, velocity and velocity dispersion degrees of freedom is investigated in a non-perturbative approximation scheme based on the Dyson–Schwinger equations. In particular, the generation of vorticity and velocity dispersion is studied and predictions for the corresponding power spectra are made, which qualitatively agree well with results obtained from $N$-body simulations. It is found that velocity dispersion grows strongly due to non-linear effects and at late times its mean value seems to be largely independent of the initial conditions. By taking this into account, a rather realistic picture of non-linear large-scale structure formation can be obtained, albeit the numerical treatment remains challenging, especially for very cold dark matter models.

## Contents

## 1 Introduction

One of the primary goals of contemporary cosmology is to describe the evolution of dark matter under the influence of gravity, in order to understand the observed large-scale structure of the Universe. As modern and future observations can probe increasingly smaller scales, there is a genuine need to understand the dynamics governing the relevant physical processes at these scales.

In kinetic theory, dark matter is described by a phase-space distribution function which evolves with Boltzmann's equation. At cosmological late times, microscopic interaction rates are negligible and the distribution function's evolution is sufficiently well described by the (collisionless) Vlasov equation [1]. On scales much smaller than the Hubble horizon, gravity around an expanding, homogeneous and isotropic background cosmology is, in leading order approximation, described by Poisson's equation. The so-called Vlasov–Poisson system of equations is a closed system of non-linear integro-differential equations in seven variables and, as such, is rather difficult to solve in full generality. From an observational point of view, one is often interested in velocity moments or cumulants instead of the full phase-space distribution function. The Vlasov–Poisson system of equations can be cast into an infinite hierarchy of coupled evolution equations for the distribution function's moments or cumulants [2].

In order to obtain a closed and finite set of equations, a common approach is to truncate the distribution function's infinite cumulant expansion after the first order. This often used single-stream approximation models dark matter as a perfect pressureless fluid and is preserved by the Vlasov–Poisson system of equations, in the sense that no higher order cumulants are sourced if initially absent. However, it is an apparent self-consistency due to the phenomenon of shell-crossing. During gravitational collapse dark matter particles meet in position space and generate a non-trivial velocity dispersion tensor which indicates the break down of the single-stream approximation. While the perfect pressureless fluid description provides a surprisingly good account of early stage gravitational instabilities, one naturally longs for a description that is capable to capture the physics of a multi-stream flow. Ultimately, this relies on including some notion of velocity dispersion. To this end, various approaches have been developed, ranging from a description including the velocity dispersion tensor [3–7], and more recently even higher-order cumulants [8, 9], through functional approaches, either over the phase-space distribution function [10] or the Lagrangian displacement field [11], to using the Schrödinger method to tackle the cumulant hierarchy beyond a truncation [12, 13], to name only a few.

In this work a truncation of the cumulant expansion after the second order is pursued, which includes the velocity dispersion tensor as dynamical degrees of freedom. Within this truncation, the single-stream approximation's degenerate distribution function is smoothened out, which allows to capture the average motion of a multi-stream flow. This should be understood in terms of an effective macroscopic description which models dark matter in a fluid dynamical approach with some notion of 'warmness'. Moreover, the inclusion of velocity dispersion is naturally paired with non-trivial dynamics for the velocity's rotational mode. Vorticity is not sourced if primordially absent in the single-stream approximation, but is known to be generated at shell-crossing [14].

A common approach to study cosmic structure formation is standard perturbation theory. Formally assuming that the deviations from a homogeneous and isotropic background cosmology are small, the dynamics are solved around linear theory and non-linear corrections are organised in an expansion in the initial correlation of mass density fluctuations. For sufficiently early times and large scales, perturbation theory captures the relevant dynamics, whereas at later times and smaller scales the perturbative series badly fails to converge since the deviations from the background become large. In order to access the non-linear regime various approaches have been proposed, including partial resummations of the perturbation series [15–25], effective theories [26–29] or renormalisation group approaches [30–35]. To this end, this work studies a non-perturbative approach based on the Dyson–Schwinger equations of the underlying statistical field theory. An approximation involving cosmic one- and two-point correlation functions is studied for the cosmological fluid. However, this approach can easily be systematically extended as is done for example in quantum field theories.

The paper is structured as follows. In Section 2 the Vlasov–Poisson system of equations is presented and the distribution function's cumulant expansion reviewed. Section 3 revises the Martin–Siggia–Rose/Janssen–de Dominicis formalism for Langevin type dynamics with Gaussian random field initial conditions. The corresponding Dyson–Schwinger equations are derived and truncated in order to obtain a closed set of evolution equations for one- and two-point correlation functions. Section 4 utilises the truncated Dyson–Schwinger equations in a variant of Hartree's approximation to evolve cosmic correlation functions. This is extended to a self-consistent Hartree–Fock approximation in Section 5, an approach which features genuinely non-perturbative physics. Some conclusions are drawn in Section 6. Appendix A provides details about the employed stochastic response field formalism and Appendix B collects explicit expressions for bare vertices.

## 2 Gravitational dynamics of dark matter and statistical description

The gravitational dynamics of an ensemble of non-relativistic and collisionless point particles of mass $m$ is described by the Vlasov–Poisson system of equations [1],

$$\begin{aligned} \partial_\tau f + \frac{p_i}{am}\,\partial_i f - am\,\partial_i\phi\,\frac{\partial f}{\partial p_i} &= 0\;, \\ \partial_i\partial_i\phi = \tfrac{3}{2}\mathcal{H}^2\Omega_\mathrm{m}&\left[\int_{\mathbb{R}^3}\frac{\mathrm{d}^3p}{(2\pi)^3}\,f - 1\right]. \end{aligned} \tag{2.1}$$

Here, $\tau$ is conformal time, $\boldsymbol{x}$ are comoving coordinates and $\boldsymbol{p} = am\,\mathrm{d}\boldsymbol{x}/\mathrm{d}\tau$ are the corresponding conjugate momenta.[1] The former two are related to cosmic time and proper coordinates by $\mathrm{d}t = a(\tau)\,\mathrm{d}\tau$ and $\boldsymbol{r} = a(\tau)\,\boldsymbol{x}$, respectively, where $a(\tau)$ is the scale factor, parametrising the expansion of the spatial Universe. Further, $\mathcal{H} = \mathrm{d}a/\mathrm{d}t$ is the conformal Hubble rate and $\Omega_\mathrm{m}$ is the (time-dependent) dark matter density parameter. Finally, $f(\tau, \boldsymbol{x}, \boldsymbol{p})$ is the dark matter one-particle phase-space distribution function, here normalised to

$$\int_{\mathbb{R}^3}\frac{\mathrm{d}^3p}{(2\pi)^3}\,\langle f(\tau,\boldsymbol{x},\boldsymbol{p})\rangle = 1\;, \tag{2.2}$$

and $\phi(\tau, \boldsymbol{x})$ is the Newtonian gravitational potential.[2]

When studying late-time cosmic structure formation, one is often less interested in the full phase-space distribution function but rather in velocity *moments* or *cumulants* thereof, the full set of which completely characterises $f(\tau, \boldsymbol{x}, \boldsymbol{p})$. The first few velocity moments are given by the (background-normalised) dark matter mass density,

$$1 + \delta = \int_{\mathbb{R}^3}\frac{\mathrm{d}^3p}{(2\pi)^3}\,f\;, \tag{2.3}$$

the momentum density,

$$(1+\delta)\,u_i = \int_{\mathbb{R}^3}\frac{\mathrm{d}^3p}{(2\pi)^3}\,\frac{p_i}{am}\,f\;, \tag{2.4}$$

and the stress tensor,

$$(1+\delta)(u_iu_j + \sigma_{ij}) = \int_{\mathbb{R}^3}\frac{\mathrm{d}^3p}{(2\pi)^3}\,\frac{p_i}{am}\frac{p_j}{am}\,f\;, \tag{2.5}$$

here expressed in terms of the density contrast field $\delta(\tau, \boldsymbol{x})$, quantifying the local deviation from the mean dark matter mass density, the velocity vector field $u_i(\tau, \boldsymbol{x})$ and the velocity dispersion tensor field $\sigma_{ij}(\tau, \boldsymbol{x})$. The latter two are the first- and second-order velocity cumulants of the distribution function, respectively.

[1] Symbols with indices from the middle of the Latin alphabet refer to spatial vector components, while boldface symbols denote the corresponding vector. Indices appearing twice in a single term are summed over, where vectors and covectors are not distinguished and are both denoted with subscripted indices, as is common in flat space. Partial derivatives with respect to the $i$th comoving coordinate component are abbreviated by $\partial_i$, while all other partial derivative are denoted by a suitably subscripted $\partial$. Finally, function arguments are often suppressed for the sake of brevity.

[2] The expectation value $\langle\,\cdot\,\rangle$ is either taken as an ensemble average over cosmic histories with stochastic initial conditions or as a sample average over large spatial volumes in a single cosmic history.

Instead of dealing with the momentum dependence of the distribution function, the Vlasov–Poisson system of equations (2.1) can be cast into an infinite tower of coupled evolution equations for the velocity moments or cumulants [2]. For practical purposes, one often turns to approximations describing dark matter by a finite number of velocity cumulants to obtain a closed system of equations. The most prominent example is the so-called *single-stream approximation*, which models dark matter as a perfect pressureless fluid that is described by the density contrast field $\delta$ and velocity field $u_i$ only [36].

Although the single-stream approximation successfully describes the gravitational dynamics of dark matter at early times and large scales, it fails to do so at later times and smaller scales. Most notably, the single-stream approximation cannot account for the phenomenon of *shell-crossing*, when particle trajectories meet in position space, since higher-order velocity cumulants are naturally generated, which indicates the break-down of the single-stream approximation [14].

In the following, the description of dark matter is extended beyond the single-stream approximation by including the velocity dispersion tensor $\sigma_{ij}$ as a dynamical degree of freedom to describe gravitational dynamics also after shell-crossing. On the level of the phase-space distribution function, this corresponds to smoothening out the degenerate single-stream approximation distribution function to a normal distribution, where the velocity dispersion tensor is the covariance matrix of the spatial momenta [7, 35]. Naturally, this approximation cannot describe shell-crossing microscopically, but it supports the average motion of a multi-stream fluid [3] and captures signatures associated to shell-crossing, as is shown *a posteriori*. While a truncation of the cumulant expansion after the second-order is not self-consistent, in the sense that higher-order velocity cumulants are dynamically generated even if initially absent [14], one can argue that this approximation could provide a viable description of virialised clumps of dark matter, which seem to be near to a non-relativistic thermal state, at least for simple halo models [37].

In the following, higher-order velocity cumulants are neglected *cum grano salis* and dark matter is described in terms of an effective theory with ten matter degrees of freedom. These are parametrised in terms of the density contrast field $\delta$, the velocity vector field $u_i$ and the (symmetric) velocity dispersion tensor field $\sigma_{ij}$. The relevant evolution equations are given by the continuity equation,

$$\partial_\tau \delta + \partial_i[(1+\delta)u_i] = 0 \,, \tag{2.6}$$

the Cauchy momentum equation,

$$\partial_\tau u_i + \mathcal{H} u_i + \partial_j \sigma_{ij} + u_j \partial_j u_i + \sigma_{ij} \partial_j \ln(1+\delta) + \partial_i \phi = 0 \,, \tag{2.7}$$

and the velocity dispersion equation,

$$\partial_\tau \sigma_{ij} + 2\mathcal{H}\sigma_{ij} + u_k \partial_k \sigma_{ij} + \sigma_{jk}\partial_k u_i + \sigma_{ik}\partial_k u_j = 0 \,, \tag{2.8}$$

which together with Poisson's equation,

$$\partial_i \partial_i \phi = \tfrac{3}{2}\mathcal{H}^2 \Omega_\mathrm{m} \delta \,, \tag{2.9}$$

form a closed system of equations.[3]

[3]While equations (2.6) and (2.7) are exact, the terms which involve the third-order velocity cumulant in equation (2.8) are neglected, thereby closing the system of equations.

In cosmology it is common to adopt a statistical description for dark matter, where the relevant degrees of freedom are described in terms of random fields. This is most often motivated by a very early inflationary epoch of the Universe, which predicts the initial state of the theory to be very well described by Gaussian random fields that are statistically homogeneous and isotropic in space [38]. The *fair sample hypothesis* asserts that the employed statistical field theory can be understood as describing an ensemble of cosmic histories with stochastic initial conditions or a sample of large spatial volumes within a single cosmic history [39].

To investigate the statistical properties of cosmic large-scale structures, one is interested in *correlation functions* of the velocity moments and cumulants, or more generally the phase-space distribution function. Since the underlying random fields are homogeneous and isotropic in space, it is useful to decompose the matter degrees of freedom with respect to spatial translations and rotations, which is most conveniently done using the Fourier transform.[4] For the velocity vector field the decomposition can be parametrised as

$$u_i(\boldsymbol{x}) = \int_{\boldsymbol{k}} \mathrm{e}^{\mathrm{i}\boldsymbol{k}\cdot\boldsymbol{x}} \left[ \epsilon_{ijl} \, \frac{\mathrm{i}\, k_j}{k^2} \, \omega_l(\boldsymbol{k}) - \frac{\mathrm{i}\, k_i}{k^2} \, \theta(\boldsymbol{k}) \right] , \tag{2.10}$$

where i is the imaginary unit and $\epsilon_{ijk}$ is the totally antisymmetric Levi-Civita symbol. The scalar velocity divergence mode $\theta(\boldsymbol{k})$ parametrises the conservative part of the velocity field, while the pseudovector vorticity mode $\omega_i(\boldsymbol{k})$ quantifies the solenoidal part. Similarly, the velocity dispersion tensor field can be decomposed as

$$\sigma_{ij}(\boldsymbol{x}) = \int_{\boldsymbol{k}} \mathrm{e}^{\mathrm{i}\boldsymbol{k}\cdot\boldsymbol{x}} \left[ \vartheta_{ij}(\boldsymbol{k}) + \hat{k}_i \vartheta_j(\boldsymbol{k}) + \hat{k}_j \vartheta_i(\boldsymbol{k}) + \tfrac{3}{2} \left( \hat{k}_i \hat{k}_j - \tfrac{1}{3}\, \delta_{ij} \right) \vartheta(\boldsymbol{k}) + \delta_{ij} \sigma(\boldsymbol{k}) \right] , \tag{2.11}$$

where $\hat{k}_i = k_i/k$ and $\delta_{ij}$ is the Kronecker delta. The scalar trace mode $\sigma(\boldsymbol{k})$ parametrises isotropic velocity dispersion, while the scalar mode $\vartheta(\boldsymbol{k})$, the transverse vector mode $\vartheta_i(\boldsymbol{k})$ and the transverse traceless tensor mode $\vartheta_{ij}(\boldsymbol{k})$ parametrise anisotropic velocity dispersion degrees of freedom. The decompositions (2.10) and (2.11) correspond to splitting the velocity and velocity dispersion degrees of freedom $3 \to 2 + 1$ and $6 \to 2 + 2 + 1 + 1$, respectively.

Due to the statistical symmetries, the one-point correlation functions (*mean fields*) are given by

$$\langle \delta(\tau, \boldsymbol{x}) \rangle = 0 \, , \qquad \langle u_i(\tau, \boldsymbol{x}) \rangle = 0 \, , \qquad \langle \sigma_{ij}(\tau, \boldsymbol{x}) \rangle = \delta_{ij} \, \bar{\sigma}(\tau) \, , \tag{2.12}$$

[4]The splitting into Fourier modes is a decomposition into irreducible unitary representations of the translation group and for any suitable function or distribution $h(\boldsymbol{x})$ the convention

$$h(\boldsymbol{k}) = \int_{\boldsymbol{x}} \mathrm{e}^{-\mathrm{i}\boldsymbol{k}\cdot\boldsymbol{x}} \, h(\boldsymbol{x}) \, , \qquad h(\boldsymbol{x}) = \int_{\boldsymbol{k}} \mathrm{e}^{\mathrm{i}\boldsymbol{k}\cdot\boldsymbol{x}} \, h(\boldsymbol{k}) \, ,$$

is employed, where integrals are abbreviated as

$$\int_{\boldsymbol{x}} = \int_{\mathbb{R}^3} \mathrm{d}^3 x \, , \qquad \int_{\boldsymbol{k}} = \int_{\mathbb{R}^3} \frac{\mathrm{d}^3 k}{(2\pi)^3} \, .$$

The Fourier transform is denoted by the same symbol and distinguished by its argument if not clear from context. The three-dimensional Euclidean inner product is written as $\boldsymbol{k} \cdot \boldsymbol{x} = k_i x_i$ and the modulus of three-vectors is denoted by the corresponding lightface symbol, such as the wave number $k = |\boldsymbol{k}|$. Finally, wave vector Dirac delta functions are often denoted by $\bar{\delta}(\boldsymbol{k}) = (2\pi)^3 \, \delta^{(3)}(\boldsymbol{k})$.

where $\bar{\sigma}$ is a (time-dependent) isotropic velocity dispersion background.[5] For thermally produced dark matter $\bar{\sigma}$ is related to the particle mass and decoupling temperature, but generally depends on the type of candidate considered [40].

Two-point correlation functions (*covariance functions*) are of natural importance in cosmology due to the Gaussian nature of the initial state. In terms of the Fourier transform these can be quantified by their power spectral density, e.g. the matter power spectrum,

$$\langle \delta(\tau, \boldsymbol{k}) \, \delta(\tau', \boldsymbol{k}') \rangle = \delta\!\!\!^-(\boldsymbol{k} + \boldsymbol{k}') \, P_{\delta\delta}(\tau, \tau', k) \; . \tag{2.13}$$

Due to the statistical symmetries, only covariance functions of the same type of modes (scalar, transverse vector or transverse traceless tensor) are non-vanishing. As such, the transverse vector mode power spectra carry an additional projector $\mathrm{P}_{ij}(\boldsymbol{k}) = \delta_{ij} - \hat{k}_i \hat{k}_j$, e.g. the vorticity power spectrum,

$$\langle \omega_i(\tau, \boldsymbol{k}) \, \omega_j(\tau', \boldsymbol{k}') \rangle = \delta\!\!\!^-(\boldsymbol{k} + \boldsymbol{k}') \, \mathrm{P}_{ij}(\boldsymbol{k}) \, P_{\omega\omega}(\tau, \tau', k) \; , \tag{2.14}$$

and similar for the transverse traceless tensor mode power spectra.

Although late-time cosmic structure formation is naturally non-Gaussian since the evolution equations are non-linear, the power spectra are usually the objects of main interest in cosmology. However, higher-order correlation functions, such as the bi- and trispectrum, are also frequently studied and provide insights into deviations from Gaussian statistics of dark matter.

# 3 Cosmological field theory

## 3.1 Functional approach

To study correlation functions, such as mean fields and covariance functions, it is convenient to work in a field-theoretic functional formulation [41–43], some aspects of which are recapitulated in Appendix A.

The focus of this work is to investigate the evolution of the four scalar and the vorticity vector mode, while the velocity dispersion vector and tensor modes are neglected here.[6] The decomposed degrees of freedom are assembled in the field multiplet

$$\psi_a(\eta, \boldsymbol{k}) = \left( \delta \, , \, -\frac{\theta}{f_+ \mathcal{H}} \, , \, \frac{k^2 \sigma}{(f_+ \mathcal{H})^2} \, , \, \frac{k^2 \vartheta}{(f_+ \mathcal{H})^2} \, , \, \frac{\omega_i}{f_+ \mathcal{H}} \right) \; , \tag{3.1}$$

where the index $a$ runs through the field content, is summed over if appearing twice in a single term and carries any additional tensorial substructure of the fields. In the following, it is useful to work in terms of the time evolution parameter $\eta = \ln(D_+)$, where $D_+$ is the fastest growing linear density contrast mode in the single-stream approximation.[7] Further, it is sensible to introduce $f_+ = \mathrm{d}\ln(D_+)\,/\,\mathrm{d}\ln(a)$, quantifying the deviation of $D_+$ from the EdS linear growth function.[8] Finally, it is convenient to introduce the rescaled mean field $\tilde{\sigma} = \bar{\sigma}/(f_+\mathcal{H})^2$, such that $\langle\psi_a(\eta,\boldsymbol{k})\rangle = \delta_{a3}\,\bar{\delta}(\boldsymbol{k})\,k^2\tilde{\sigma}(\eta)$.[9]

[5] The velocity dispersion mean field naturally feeds into Friedmann's equations that govern the evolution of the scale factor $a(\tau)$, but the corresponding contribution is of the order $\bar{\sigma}/c^2$, with $c$ the speed of light, and is thus negligible for cold enough dark matter.

[6] The inclusion of velocity dispersion degrees of freedom allows to source the vorticity field non-linearly, whereas the vorticity field decays linearly in the single-stream approximation due to the Hubble expansion and cannot be sourced if initially absent [36]. The velocity dispersion vector and tensor degrees of freedom are neglected here due to computational limitations. Their impact is expected to be subdominant compared to the scalar velocity dispersion degrees of freedom, as the vector and tensor modes can only be sourced non-linearly (if initially absent) from scalar modes due to the structure of the non-linear terms in the velocity dispersion equation (2.8).

In terms of the field multiplet (3.1), the bare equations of motion (2.6), (2.7) and (2.8) can be written as

$$\left[\partial_\eta\delta_{ab} + \Omega_{ab}(\eta)\right]\psi_b(\eta,\boldsymbol{k}) + I_a(\eta,\boldsymbol{k}) = 0\;, \tag{3.2}$$

where the linear dynamics is specified by the matrix

$$\Omega_{ab}(\eta) = \begin{pmatrix} 0 & -1 & 0 & 0 & 0 \\ -\frac{3}{2}\frac{\Omega_\mathrm{m}}{f_+^2} & \frac{3}{2}\frac{\Omega_\mathrm{m}}{f_+^2}-1 & 1 & 1 & 0 \\ 0 & 0 & 3\,\frac{\Omega_\mathrm{m}}{f_+^2}-2 & 0 & 0 \\ 0 & 0 & 0 & 3\,\frac{\Omega_\mathrm{m}}{f_+^2}-2 & 0 \\ 0 & 0 & 0 & 0 & \frac{3}{2}\frac{\Omega_\mathrm{m}}{f_+^2}-1 \end{pmatrix}\;, \tag{3.3}$$

after eliminating the gravitational potential using Poisson's equation (2.9). Here and in the following, the vector mode sector of objects with a matrix structure in field space is implicitly understood to carry an appropriate transverse projector. The non-linear dynamics are characterised by $I_a$, which can be formally written in a vertex expansion as

$$I_a(\eta,\boldsymbol{k}) = \int_{\boldsymbol{k}'}\gamma_{abc}(\boldsymbol{k}',\boldsymbol{k}-\boldsymbol{k}')\,\psi_b(\eta,\boldsymbol{k}')\,\psi_c(\eta,\boldsymbol{k}-\boldsymbol{k}') + \mathcal{O}(\psi^3)\;, \tag{3.4}$$

with the (symmetric) vertices $\gamma_{abc}(\boldsymbol{k}_1,\boldsymbol{k}_2) = \gamma_{acb}(\boldsymbol{k}_2,\boldsymbol{k}_1)$, some of which are explicitly given in Appendix B.[10]

In the context of cosmology, one is interested in expectation values of composite fields of the field content (3.1) which obey the equations of motion (3.2) for Gaussian random initial conditions. As such, the initial state is completely characterised by the initial velocity dispersion mean field $\tilde{\sigma}^\mathrm{in}$ and the power spectra $P^\mathrm{in}_{ab}(k)$. Due to statistical isotropy and homogeneity, the mean field is a zero mode and the power spectra depend on one wave number only, up to possible projectors for vector and tensor modes. Since the bare equations of motion (3.2) are non-linear, the fields naturally evolve away from their initial Gaussian shape and are completely characterised by an infinite set of correlation functions.

---

[7] For a Lambda cold dark matter ($\Lambda$CDM) cosmology without a radiative component, as applicable for late-time cosmic structure formation deep within the matter dominated era of the Universe, the linear growth function is given by [44]

$$D_+ = a\;{}_2F_1\!\left(\tfrac{1}{3},\,1;\,\tfrac{11}{6};\,-\varpi\,a^3\right)\Big/\;{}_2F_1\!\left(\tfrac{1}{3},\,1;\,\tfrac{11}{6};\,-\varpi\right)\,,$$

where ${}_2F_1$ is the Gaussian hypergeometric function and $\varpi = 1/\Omega_\mathrm{m}(\tau_0) - 1$, with $\tau_0$ the conformal time corresponding to today. Here, the growth function is normalised to $D_+(\tau_0) = 1$ and in the Einstein–de Sitter (EdS) limit $\Omega_\mathrm{m}\to 1$ one obtains $D_+ = a$.

[8] The approximation $\Omega_\mathrm{m}/f_+^2 = 1$ allows to map a $\Lambda$CDM to an EdS cosmology, where the linear equations of motion are time-translation invariant and often admit analytical solutions.

[9] Since the physical part of the mean field is $\tilde{\sigma}$, zero modes need to be treated cautiously and expressions of the form $\bar{\delta}(\boldsymbol{k})\,k^2\,F(k,q)$ are only to be evaluated under an integral.

[10] The higher-order terms of the vertex expansion are due to the term $\sigma_{ij}\partial_j\ln(1+\delta)$ appearing in equation (2.7) and are formally only justified for $\delta < 1$. In a field-theoretic treatment, the higher-order vertices involve higher loop orders which is beyond the approximations investigated in Section 3.2.

In the functional formulation the system is characterised by the bare action

$$\begin{aligned} S[\psi,\hat{\psi}] = &-\mathrm{i}\int_{\eta,\boldsymbol{k}} \hat{\psi}_a(\eta,-\boldsymbol{k})\,\Big[\big[\partial_\eta\delta_{ab}+\Omega_{ab}(\eta)\big]\psi_b(\eta,\boldsymbol{k})+I_a(\eta,\boldsymbol{k})\Big] \\ &+\int_{\boldsymbol{k}} \hat{\psi}_a(\eta_{\rm in},-\boldsymbol{k})\Big[\mathrm{i}\,\delta(\boldsymbol{k})\,\delta_{a3}\,k^2\tilde{\sigma}^{\rm in}+\tfrac{1}{2}P^{\rm in}_{ab}(k)\,\hat{\psi}_b(\eta_{\rm in},\boldsymbol{k})\Big]\ , \end{aligned} \tag{3.5}$$

where the first line characterise the (bare) dynamics, while the second line entails the Gaussian statistics of the initial state.

From the generating functional of connected correlation functions $W$, one obtains the mean fields at vanishing source currents from

$$W^{(1)}_{\boldsymbol{A}} = \delta(\boldsymbol{k})\begin{pmatrix}\delta_{a3}\,k^2\tilde{\sigma}(\eta)\\ 0\end{pmatrix}\ , \tag{3.6}$$

while connected two-point functions are given by

$$W^{(2)}_{\boldsymbol{AB}} = \delta(\boldsymbol{k}+\boldsymbol{k}')\begin{pmatrix}P_{ab}(\eta,\eta',k) & \mathrm{i}\,G^{\rm R}_{ab}(\eta,\eta',k)\\ \mathrm{i}\,G^{\rm A}_{ab}(\eta,\eta',k) & 0\end{pmatrix}\ . \tag{3.7}$$

Here, $P_{ab}(\eta,\eta',k)$ is the power spectrum of two fields and $G^{\rm R}_{ab}(\eta,\eta',k)$ is the (retarded) mean linear response function, in cosmology often called propagator, to which the advanced counterpart is related by $G^{\rm A}_{ab}(\eta,\eta',k)=G^{\rm R}_{ba}(\eta',\eta,k)$.[11]

In the following, it is convenient to also work with the one-particle irreducible (1PI) effective action $\Gamma$, which is the generating functional of 1PI correlation functions. At vanishing source currents, the 1PI one-point functions are

$$\Gamma^{(1)}_{\boldsymbol{A}} = \delta(\boldsymbol{k})\begin{pmatrix}0\\ -\mathrm{i}\,\delta_{a3}\,k^2 E(\eta)\end{pmatrix}\ , \tag{3.8}$$

while the 1PI two-point functions take the form

$$\Gamma^{(2)}_{\boldsymbol{AB}} = \delta(\boldsymbol{k}+\boldsymbol{k}')\begin{pmatrix}0 & -\mathrm{i}\,D^{\rm A}_{ab}(\eta,\eta',k)\\ -\mathrm{i}\,D^{\rm R}_{ab}(\eta,\eta',k) & H_{ab}(\eta,\eta',k)\end{pmatrix}\ . \tag{3.9}$$

Here, $E(\eta)$ is the effective equation of motion for the mean field $\tilde{\sigma}(\eta)$, $D^{\rm A}_{ab}(\eta,\eta',k)$ and $D^{\rm R}_{ab}(\eta,\eta',k)$ are the inverse advanced and retarded propagators, respectively, and $H_{ab}(\eta,\eta',k)$ is the statistical 1PI two-point correlation function.

At vanishing source currents, the effective equation of motion reads

$$E(\eta) = 0\ , \tag{3.10}$$

the propagator equation is

$$\int_\xi D^{\rm R}_{ab}(\eta,\xi,k)\,G^{\rm R}_{bc}(\xi,\eta',k) = \delta_{ac}\delta(\eta-\eta')\ , \tag{3.11}$$

[11] The retarded propagator is causal, i.e. vanishes for $\eta<\eta'$, and obeys $G^{\rm R}_{ab}(\eta,\eta',k)\to\delta_{ab}\theta(0)$ in the limit from below $\eta'\to\eta^-$. Here, the value of the Heaviside unit step function at vanishing argument depends on the employed discretisation convention for stochastic differential equations and is in the following set to $\theta(0)=0$, corresponding to Itô's calculus.

and the power spectrum equation is

$$\int_\xi D^{\mathrm{R}}_{a\bar{a}}(\eta,\xi,k)\,P_{\bar{a}b}(\xi,\eta',k) = \int_{\xi'} H_{a\bar{b}}(\eta,\xi',k)\,G^{\mathrm{A}}_{\bar{b}b}(\xi',\eta',k)\;. \tag{3.12}$$

Equations (3.10), (3.11) and (3.12) are exact and although they cannot be straightforwardly solved since the 1PI one- and two-point functions are not explicitly known, they provide a generically non-perturbative way to compute the mean field and connected two-point correlation functions.

To investigate approximation schemes, it is sensible to split the 1PI one- and two-point functions into a linear contribution and a non-linear correction, where the former can be directly read off the bare action (3.5) and the latter parametrises the ignorance of the 1PI functions. For the effective equation of motion of the velocity dispersion mean field this amounts to

$$E(\eta) = \partial_\eta\tilde{\sigma}(\eta) + (3\Omega_{\mathrm{m}}/f_+^2 - 2)\,\tilde{\sigma}(\eta) - \delta(\eta-\eta_{\mathrm{in}})\,\tilde{\sigma}^{\mathrm{in}} - Q(\eta)\;, \tag{3.13}$$

where the second and third terms on the right-hand side are the Hubble drag term and the initial condition, respectively. The non-linear correction is parametrised in terms of the *source* $Q(\eta)$, which quantifies the back-reaction of fluctuations onto the mean field. Similarly, the 1PI two-point functions (3.9) can be decomposed as

$$\begin{aligned} D^{\mathrm{R}}_{ab}(\eta,\eta',k) &= \big[\partial_\eta\delta_{ab} + \Omega_{ab}(\eta)\big]\,\delta(\eta-\eta') - \Sigma^{\mathrm{R}}_{ab}(\eta,\eta',k)\;,\\ H_{ab}(\eta,\eta',k) &= \delta(\eta-\eta_{\mathrm{in}})\,\delta(\eta'-\eta_{\mathrm{in}})\,P^{\mathrm{in}}_{ab}(k) + \Pi_{ab}(\eta,\eta',k)\;, \end{aligned} \tag{3.14}$$

where the non-linear corrections are parametrised in terms of the *self-energies* $\Sigma^{\mathrm{R}}_{ab}(\eta,\eta',k)$ and $\Pi_{ab}(\eta,\eta',k)$. Further, it is convenient to split the inverse propagator self-energy into parts that are local and non-local in time,

$$\Sigma^{\mathrm{R}}_{ab}(\eta,\eta',k) = \delta(\eta-\eta')\,\Sigma^{\mathrm{H}}_{ab}(\eta,k) + \Sigma^{\mathrm{F}}_{ab}(\eta,\eta',k)\;, \tag{3.15}$$

which in the following are referred to as *Hartree* and *Fock self-energies*, respectively. Since the statistical self-energy $\Pi_{ab}(\eta,\eta',k)$ is naturally non-local in time, it is also regarded as a Fock-type self-energy.

Eliminating the advanced in favour of the retarded propagator and plugging the expressions (3.13), (3.14) and (3.15) into the evolution equations (3.10), (3.11) and (3.12), one obtains the effective equation of motion

$$\partial_\eta\tilde{\sigma}(\eta) + (3\Omega_{\mathrm{m}}/f_+^2 - 2)\,\tilde{\sigma}(\eta) = \delta(\eta-\eta_{\mathrm{in}})\,\tilde{\sigma}^{\mathrm{in}} + Q(\eta)\;, \tag{3.16}$$

the propagator equation

$$\begin{aligned} \big[\partial_\eta\delta_{ab} + \Omega_{ab}(\eta) - \Sigma^{\mathrm{H}}_{ab}(\eta,k)\big]\,G^{\mathrm{R}}_{bc}(\eta,\eta',k) - \int_{\eta'}^{\eta} \mathrm{d}\xi\,\Sigma^{\mathrm{F}}_{ab}(\eta,\xi,k)\,G^{\mathrm{R}}_{bc}(\xi,\eta',k)&\\ = \delta_{ac}\delta(\eta-\eta')\;,& \end{aligned} \tag{3.17}$$

and the power spectrum equation

$$\begin{aligned} \big[\partial_\eta\delta_{a\bar{a}} + \Omega_{a\bar{a}}(\eta) - \Sigma^{\mathrm{H}}_{a\bar{a}}(\eta,k)\big]\,P_{\bar{a}b}(\eta,\eta',k) - \int_{\eta_{\mathrm{in}}}^{\eta} \mathrm{d}\xi\,\Sigma^{\mathrm{F}}_{a\bar{a}}(\eta,\xi,k)\,P_{\bar{a}b}(\xi,\eta',k)&\\ = \delta(\eta-\eta_{\mathrm{in}})\,G^{\mathrm{R}}_{b\bar{b}}(\eta',\eta_{\mathrm{in}},k)\,P^{\mathrm{in}}_{a\bar{b}}(k) + \int_{\eta_{\mathrm{in}}}^{\eta'} \mathrm{d}\xi'\,G^{\mathrm{R}}_{b\bar{b}}(\eta',\xi',k)\,\Pi_{a\bar{b}}(\eta,\xi',k)\;.& \end{aligned} \tag{3.18}$$

Up to here, no approximations have been made and equations (3.16), (3.17) and (3.18) are exact. The ignorance of the 1PI one- and two-point functions has simply been absorbed into the source $Q$ and the self-energies $\Sigma^{\mathrm{H}}_{ab}$, $\Sigma^{\mathrm{F}}_{ab}$ and $\Pi_{ab}$. Given explicit expressions for these, one can solve the evolution equations to obtain the mean field and connected two-point correlation functions. However, in general it is not possible to provide exact and closed expressions for the source and self-energies since these typically involve higher-order correlation functions. Nonetheless, the expansion in terms of connected and 1PI correlation functions naturally allows for non-perturbative investigations with functional methods, such as the Dyson–Schwinger equations [19, 42, 43] and the renormalisation group [31–35], the former of which are studied in the next section.

### 3.2 Dyson–Schwinger equations

To obtain explicit expressions for the source and self-energies, the *Dyson–Schwinger equations* [45–47] of the statistical field theory are investigated. In the current context it reads

$$\Gamma^{(1)}_{\boldsymbol{A}}[\Psi] = S^{(1)}_{\boldsymbol{A}}\left[\Psi + \left[\Gamma^{(2)}[\Psi]\right]^{-1}\cdot\frac{\delta}{\delta\Psi}\right] , \tag{3.19}$$

where the dot product runs over all internal structures, namely the physical-response field content, time and space arguments.

The Dyson–Schwinger equation (3.19) is an exact equation, which relates the 1PI one-point correlation function to its bare counterpart, i.e. the first functional derivative of the bare action, evaluated on the field configuration given in the argument on the right-hand side. Equation (3.19) can be derived from the invariance of the functional integral measure with respect to additive shifts of the fields. For the type of bare action (3.5) the Dyson–Schwinger equation (3.19) in general involves all functional derivatives of the bare and effective action.[12] This is due to the fact that the term $\sigma_{ij}\ln(1+\delta)$ involves vertices of all orders in a vertex expansion. In the following, only the bare three-point vertex is taken into account and the terms originating from higher-order vertices are neglected. In this case the Dyson–Schwinger equation (3.19) depends on the third-order functional derivative of the bare action $S^{(3)}$, that is up to normalisation the bare three-point vertex $\gamma_{abc}$, as well as the second-order functional derivative of the effective action $\Gamma^{(2)}$.

By applying functional derivatives to the Dyson–Schwinger equation (3.19) one obtains equations for higher-order 1PI correlation functions, which in turn depend on 1PI correlation functions up to the next higher order, creating an infinite hierarchy of coupled equations. More explicitly, the Dyson–Schwinger equation for the 1PI one-point function reads

$$\Gamma^{(1)}_{\boldsymbol{A}} = S^{(1)}_{\boldsymbol{A}} + \tfrac{1}{2}\,\mathrm{Tr}\left[W^{(2)}\cdot S^{(3)}_{\boldsymbol{A}}\right] , \tag{3.20}$$

while the equation for the 1PI two-point function is given by

$$\Gamma^{(2)}_{\boldsymbol{AB}} = S^{(2)}_{\boldsymbol{AB}} - \tfrac{1}{2}\,\mathrm{Tr}\left[W^{(2)}\cdot S^{(3)}_{\boldsymbol{A}}\cdot W^{(2)}\cdot \Gamma^{(3)}_{\boldsymbol{B}}\right] , \tag{3.21}$$

where the dot product and the trace run over all internal structures. The inverse of the second-order functional derivatives of the effective action is expressed in terms of the connected two-point correlation functions by virtue of equation (A.7) and should be understood as being field-dependent through the Legendre transform. Similarly, one can proceed for higher-order 1PI correlation functions to obtain an infinite tower of coupled equations.

[12] Generally, for a bare action that is maximally of polynomial degree $n$ in the fields, the Dyson–Schwinger equation (3.19) involves derivatives of the bare and effective action up to order $n$ and $(n-1)$, respectively.

The Dyson–Schwinger equations (3.20) and (3.21) can now be used to assign explicit expressions to the source and self-energies. At vanishing source currents, the mean field source reads

$$Q(\eta) = \frac{1}{6\pi^2} \int_0^\infty \mathrm{d}q \left[ 2P_{42}(\eta, \eta, q) - P_{32}(\eta, \eta, q) \right] , \tag{3.22}$$

or equivalently using the diagrammatic representations introduced in Appendix A,

$\mathrm{i}\, Q = \frac{1}{2}$ 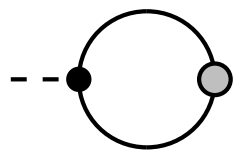 . (3.23)

From this expression it is evident that the mean field is sourced by the equal-time two-point cross-correlation of velocity and velocity dispersion fluctuations.[13]

The Hartree self-energy arises from evaluating $S^{(2)}[\Psi]$ at non-vanishing mean field and is given by

$$\Sigma^{\mathrm{H}}_{ab}(\eta, k) = \begin{pmatrix} 0 & 0 & 0 & 0 & 0 \\ -1 & 0 & 0 & 0 & 0 \\ 0 & \frac{2}{3} & 0 & 0 & 0 \\ 0 & \frac{4}{3} & 0 & 0 & 0 \\ 0 & 0 & 0 & 0 & 0 \end{pmatrix} k^2 \tilde{\sigma}(\eta) , \tag{3.24}$$

or diagrammatically as

$\mathrm{i}\, \Sigma^{\mathrm{H}}_{ab} =$ 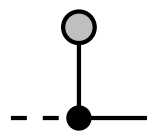 , (3.25)

and parametrises the presence of a non-vanishing mean field (velocity dispersion) onto the response of fluctuations. In contrast, the Fock self-energies contain loop expressions as parametrised by the trace term and read diagrammatically

$\mathrm{i}\, \Sigma^{\mathrm{F}}_{ab} =$ 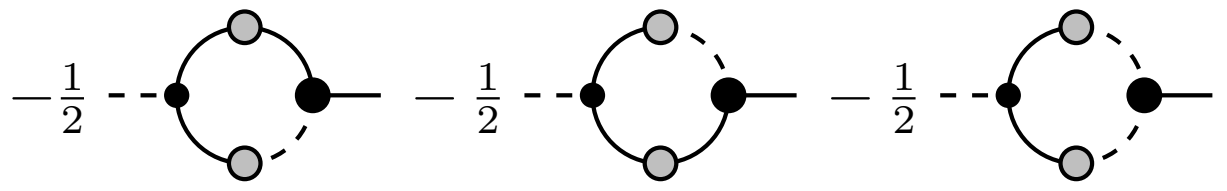 , (3.26)

and

$\Pi_{ab} =$ 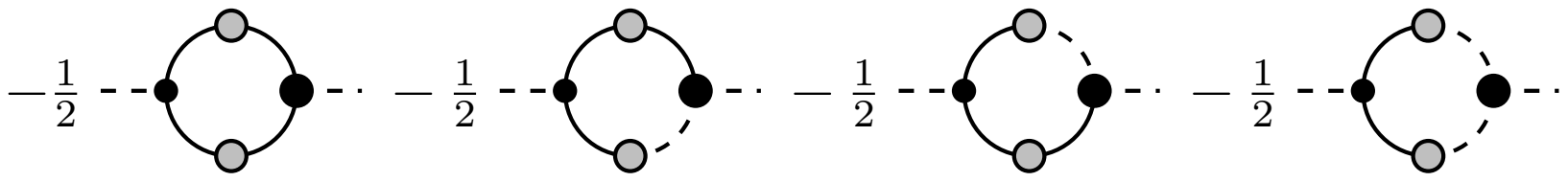 . (3.27)

The explicit formulae for the Fock self-energies are more involved and are not explicitly displayed in full form here, but are instead given later within the approximations used to close the Dyson–Schwinger hierarchy.

## 4 Hartree approximation

A simple truncation of the Dyson–Schwinger hierarchy is the Hartree approximation, where the system of equations is closed by neglecting the Fock self-energies, $\Sigma^{\mathrm{F}}_{ab} = 0$ and $\Pi_{ab} = 0$.

[13] Indeed, there is another contribution due to the cross-correlation of the velocity and velocity dispersion vector modes, which is not present in equation (3.22) due to neglecting velocity dispersion vector and tensor degrees of freedom.

The equations of motion are then modified by the presence of the time-dependent velocity dispersion mean field and the propagator equation (3.17) reads

$$\left[\partial_\eta \delta_{ab} + \varOmega_{ab}(\eta) - \varSigma^{\mathrm{H}}_{ab}(\eta,k)\right] G^{\mathrm{R}}_{bc}(\eta,\eta',k) = \delta_{ac}\delta(\eta-\eta') \,. \tag{4.1}$$

Here, the velocity dispersion mean field acts very similar to the pressure term in *Jeans instability* through the Hartree self-energy (3.24), such that growing solutions are only allowed for wave numbers smaller than the *free-streaming wave number* [7]

$$k_{\mathrm{fs}}(\eta) = \sqrt{\frac{3}{2}\frac{\varOmega_{\mathrm{m}}}{f_+^2}\frac{1}{\tilde{\sigma}}} \;. \tag{4.2}$$

The power spectrum equation (3.18) is formally solved by

$$P_{ab}(\eta,\eta',k) = G^{\mathrm{R}}_{a\bar{a}}(\eta,\eta_{\mathrm{in}},k)\, G^{\mathrm{R}}_{b\bar{b}}(\eta',\eta_{\mathrm{in}},k)\, P^{\mathrm{in}}_{\bar{a}\bar{b}}(k) \,, \tag{4.3}$$

due to the absence of the statistical self-energy. In the following, it is convenient to write the initial power spectrum as

$$P^{\mathrm{in}}_{ab}(k) = w_a(k)\, w_b(k)\, P^{\mathrm{in}}_{\delta\delta}(k) \,, \tag{4.4}$$

where $P^{\mathrm{in}}_{\delta\delta}(k)$ is the initial density contrast power spectrum and $w_1(k) = 1$. To investigate the response of fluctuations, it is convenient to define the reduced propagator

$$G^{\mathrm{R}}_a(\eta,k) = G^{\mathrm{R}}_{ab}(\eta,\eta_{\mathrm{in}},k)\, w_b(k) \,, \tag{4.5}$$

quantifying the mean linear response from initial conditions $w_a(k)$.

## 4.1 Linear mean field

To gain a better understanding of the physics captured in the Hartree approximation, it is sensible to first study a situation that can be tackled analytically. For a vanishing source term, the velocity dispersion mean field decays linearly with the Hubble expansion and one diagrammatically obtains

$$\mathrm{i}\,Q = 0 \quad , \qquad \mathrm{i}\,\varSigma^{\mathrm{H}}_{ab} =$$ 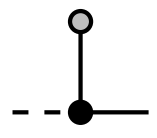 $$\quad , \qquad \mathrm{i}\,\varSigma^{\mathrm{F}}_{ab} = 0 \quad , \qquad \varPi_{ab} = 0 \quad . \tag{4.6}$$

From a physical point of view, the fluctuations evolve in the presence of a mean field, which does not feel the back-reaction of the fluctuations. Employing the approximation $\varOmega_{\mathrm{m}}/f_+^2 = 1$ and assuming initial conditions that are in the growing mode of the single-stream approximation, $w_a = (1,1,0,0,0)$, the propagator equation (4.1) can be rewritten using the reduced propagator (4.5) to arrive at a single equation for the reduced density propagator,

$$\partial_\eta^3 G^{\mathrm{R}}_1(\eta,k) + \tfrac{3}{2}\,\partial_\eta^2 G^{\mathrm{R}}_1(\eta,k) + \left[3k^2\tilde{\sigma}(\eta) - 1\right]\partial_\eta G^{\mathrm{R}}_1(\eta,k) - \tfrac{3}{2}\, G^{\mathrm{R}}_1(\eta,k) = 0 \,, \tag{4.7}$$

with the linearly decaying mean field $\tilde{\sigma}(\eta) = \tilde{\sigma}^{\mathrm{in}}\,\mathrm{e}^{-(\eta-\eta_{\mathrm{in}})}$. The general solution of this third-order differential equation can be constructed from the three independent solutions

$$\begin{aligned} F_{\mathrm{g}}(\eta,k^2\tilde{\sigma}) &= \mathrm{e}^{\eta}\,{}_1F_2\big(-1;\, -1,\, -\tfrac{3}{2};\, -3k^2\tilde{\sigma}\big) \,, \\ F_{\mathrm{d}}(\eta,k^2\tilde{\sigma}) &= \mathrm{e}^{-\frac{3}{2}\eta}\,{}_1F_2\big(\tfrac{3}{2};\, \tfrac{3}{2},\, \tfrac{7}{2};\, -3k^2\tilde{\sigma}\big) \,, \\ F^*_{\mathrm{d}}(\eta,k^2\tilde{\sigma}) &= \mathrm{e}^{-\eta}\,{}_1F_2\big(1;\, \tfrac{1}{2},\, 3;\, -3k^2\tilde{\sigma}\big) \,, \end{aligned} \tag{4.8}$$

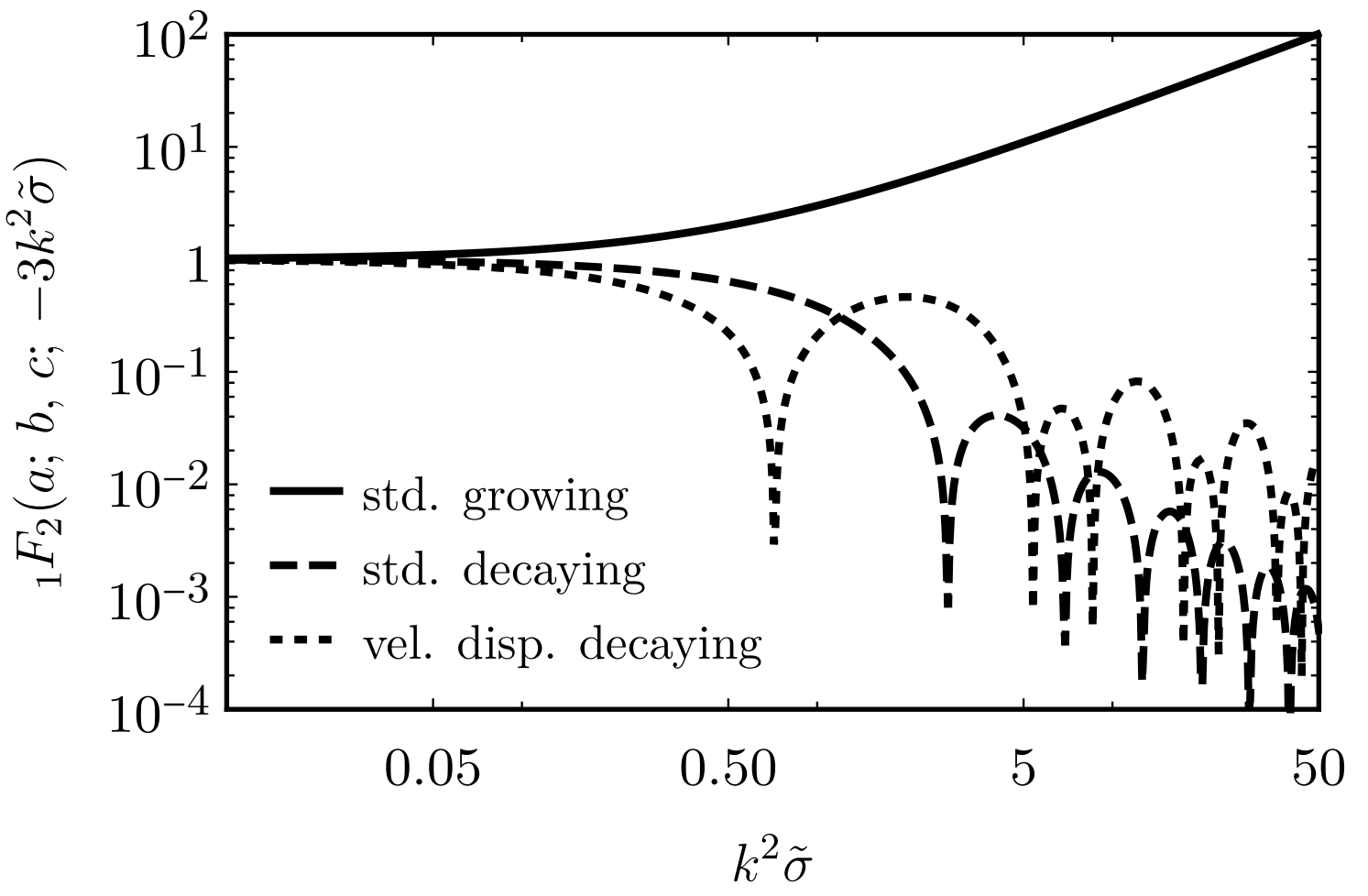

**Figure 1**. Hypergeometric functions appear in the solutions (4.8) of the reduced density propagator as a function of $k^2\tilde{\sigma}$. The growing mode (solid curve) is enhanced at non-vanishing $k^2\tilde{\sigma}$, while the decaying modes (dashed and dotted curves) are suppressed and oscillate with increasing frequency and decaying amplitude.

where ${}_1F_2\left(a; b, c; z\right)$ are generalised hypergeometric functions.

The solutions $F_\mathrm{g}$ and $F_\mathrm{d}$ smoothly connect to the standard growing and decaying modes of the single-stream approximation in the limit $k^2\tilde{\sigma} \to 0$, where the hypergeometric functions are unity, while the solution $F_\mathrm{d}^*$ is a new decaying mode due to the velocity dispersion degrees of freedom. Figure 1 displays that the growing mode is enhanced, while the decaying modes are suppressed and oscillate with increasing frequency at larger $k^2\tilde{\sigma}$.

It should be emphasised that even in this rather simple approximation, the propagator equation (3.11) is no longer time-translation invariant due to the time-dependent velocity dispersion mean field and therefore $G^\mathrm{R}_{ab}(\eta, \eta', k) \neq G^\mathrm{R}_{ab}(\eta - \eta', 0, k)$.

### 4.2 Sourced mean field

To capture the back-reaction of fluctuations onto the mean field, a non-vanishing source for the velocity dispersion mean field needs to be included. Diagrammatically this approximation reads

$\mathrm{i}\, Q = \frac{1}{2}$ 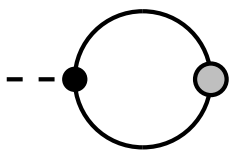 , $\mathrm{i}\, \Sigma^\mathrm{H}_{ab} =$ 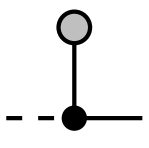 , $\mathrm{i}\, \Sigma^\mathrm{F}_{ab} = 0$ , $\Pi_{ab} = 0$ . (4.9)

That is, the velocity dispersion mean field is sourced by the back-reaction of fluctuations, which in turn evolve in the presence of the mean field. In this case, the propagator equation (3.17) has to be solved together with the mean field equation (3.16). The solution heavily depends on the initial conditions, especially on the initial velocity dispersion mean field [7, 30]. In the following, numerical solutions for different initial conditions $\tilde{\sigma}^\mathrm{in}$ are investigated, while fluctuations are initialised in the single-stream approximation growing mode, $w_a(k) = (1, 1, 0, 0, 0)$.

In Figure 2, numerical solutions of the velocity dispersion mean field for different choices of $\tilde{\sigma}^\mathrm{in}$ are shown. One finds that the smaller the initial velocity dispersion mean field is, the stronger it grows at later times. This is due to the fact that the power spectra, which source the mean field through the term (3.22), are less suppressed in the ultraviolet the smaller the value of the velocity dispersion mean field. Indeed, going to very small initial mean fields causes a rather involved numerical treatment, due to strongly suppressed and oscillating power spectra.

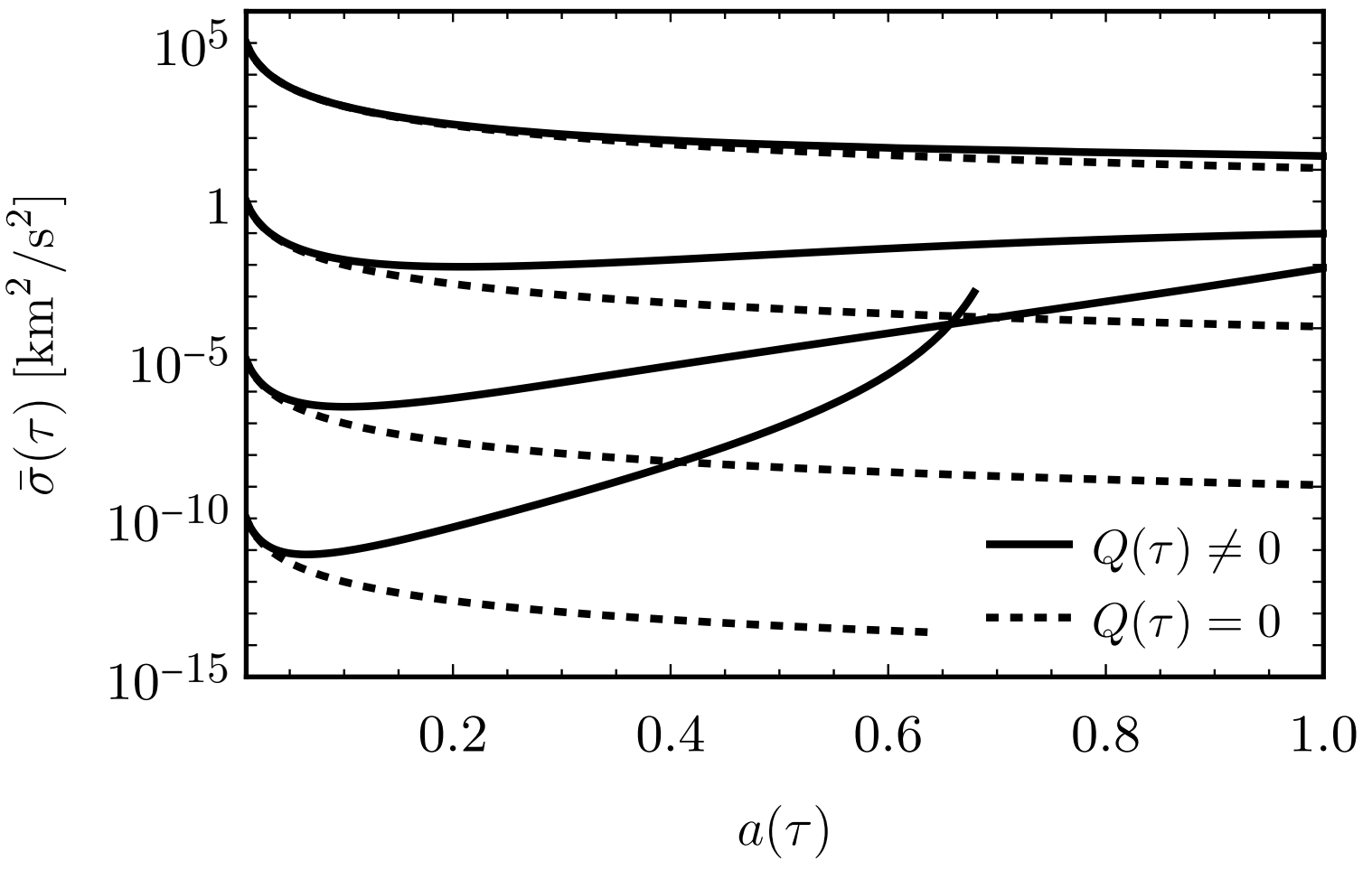

**Figure 2**. Time evolution of the velocity dispersion mean field as a function of the scale factor for different choices of initial values. Solutions which are sourced by correlations (solid curves) grow in time compared to those decaying linearly with the Hubble expansion (dotted curves). Smaller initial values of the velocity dispersion mean field correspond to larger free-streaming wave numbers and lead to a stronger late-time growth by non-linear effects.

This can be understood from the wave number dependence of the reduced propagators shown in Figure 3, which directly determine the shape of the power spectra through equation (4.4). At larger wave numbers, the propagators oscillate with increasing frequency and decaying amplitude, where the scale is set by the free-streaming wave number (4.2), very similar to the decaying solutions shown in Figure 1. The velocity dispersion-velocity cross-spectra, sourcing the mean field through the term (3.22), increasingly extend into the ultraviolet for colder dark matter models with a larger free-streaming wave number. This leads to the observed strong growth for the colder candidates displayed in Figure 2.

The drawback of this approximation is that the source of the mean field is sensitive to the ultraviolet of power spectra, a regime that is not well described by linear theory. To obtain a realistic source and mean field, non-linearities for the evolution of power spectra need to be taken into account with an intrinsically non-perturbative approximation.

## 5 Hartree–Fock approximation

A more sophisticated approximation is obtained by closing the Dyson–Schwinger hierarchy by setting the 1PI three-point functions to its bare form, while keeping the full 1PI two-point functions. This is also referred to as the self-consistent one-loop or Hartree–Fock approximation, in analogy to approximations of this type employed for quantum systems.

Because the bare vertex $\gamma_{abc}$ couples one response to two physical fields, only the first two diagrams of equation (3.26) and the first diagram of equation (3.27) survive in the

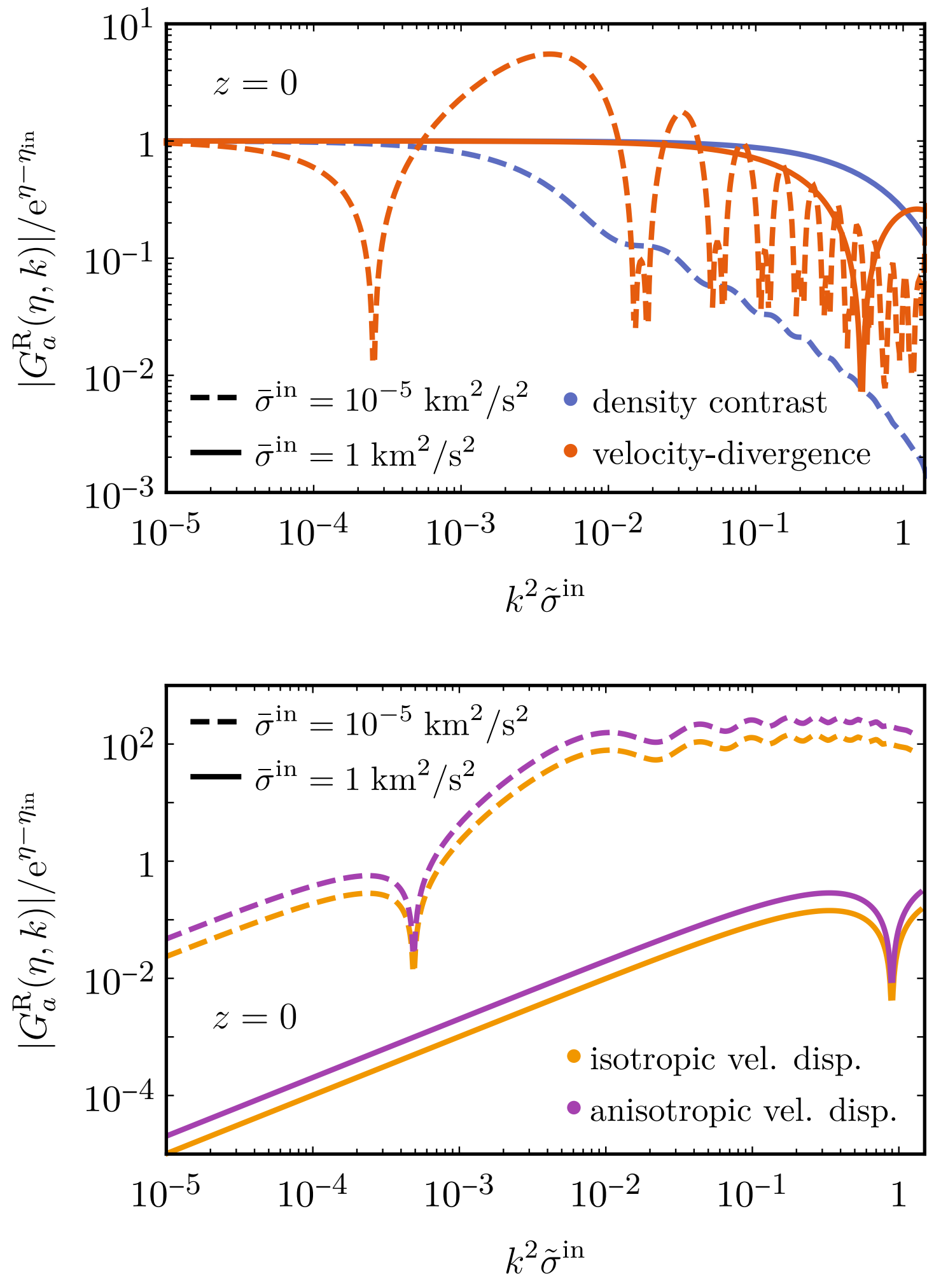

**Figure 3**. Reduced propagators for the density contrast (blue curves), velocity-divergence (red curves), isotropic velocity dispersion (yellow curves) and anisotropic velocity dispersion (violet curves) at redshift $z = 0$, normalised to the standard growing mode as a function of $k^2\tilde{\sigma}^{\text{in}}$. Results are shown for two choices of initial velocity dispersion mean fields, 1 km$^2$/s$^2$ (solid curves) and $10^{-5}$ km$^2$/s$^2$ (dashed curves). The damping scale responsible for the oscillations is set by the time-dependent mean field, shown in Figure 2, and is at much smaller $k^2\tilde{\sigma}^{\text{in}}$ for the colder dark matter model due to the strong late-time growth of the velocity dispersion mean field.

Hartree–Fock approximation. Diagrammatically one then has

$\mathrm{i}\,Q = \frac{1}{2}$ 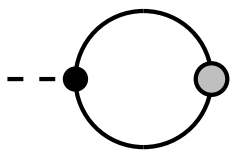 , $\mathrm{i}\,\Sigma^{\text{H}}_{ab} =$ 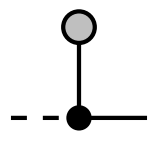 , (5.1)

and

$\mathrm{i}\,\Sigma^{\text{F}}_{ab} = -\frac{1}{2}$ 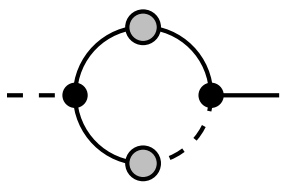 $-\frac{1}{2}$ 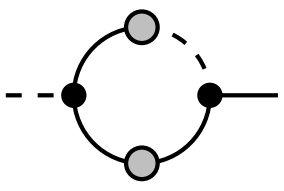 , $\Pi_{ab} = -\frac{1}{2}$ 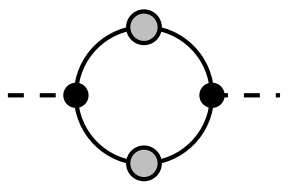 . (5.2)

More explicitly, the retarded Fock self-energy is given by

$$\Sigma^{\rm F}_{ab}(\eta,\eta',k) = 4\int_{\boldsymbol{q}} \gamma_{ace}(\boldsymbol{q},\boldsymbol{k}-\boldsymbol{q})\, G^{\rm R}_{ef}(\eta,\eta',|\boldsymbol{k}-\boldsymbol{q}|)\, \gamma_{fdb}(-\boldsymbol{q},\boldsymbol{k})\, P_{cd}(\eta,\eta',q)\ , \tag{5.3}$$

and the statistical Fock self-energy reads

$$\Pi_{ab}(\eta,\eta',k) = 2\int_{\boldsymbol{q}} \gamma_{ace}(\boldsymbol{q},\boldsymbol{k}-\boldsymbol{q})\, P_{ef}(\eta,\eta',|\boldsymbol{k}-\boldsymbol{q}|)\, \gamma_{bfd}(-\boldsymbol{k}+\boldsymbol{q},-\boldsymbol{q})\, P_{cd}(\eta,\eta',q)\ . \tag{5.4}$$

With these expressions, equations (3.16), (3.17) and (3.18) form a closed system.

### 5.1 Large external wave number limit

Before turning to full numerical results, it is instructive to consider the limit of large external wave numbers in order to get an analytical understanding of the Hartree–Fock approximation. In the limit of large external wave numbers, the propagator equation can be simplified to an extent where it can be solved in the absence of a mean field.

The inclusion of the retarded Fock self-energy $\Sigma^{\rm F}_{ab}(\eta,\eta',k)$ introduces a new scale dependence in the propagator due to the non-linear coupling of modes. As long as the velocity dispersion degrees of freedom are comparably small, one expects the dominant non-linear contribution to be due to the random advection of density structures by a large-scale velocity field, also known as the *sweeping effect* [35, 48, 49]. To isolate this contribution and investigate how it appears in the Hartree–Fock approximation, the large external wave number limit is studied in the absence of velocity dispersion degrees of freedom.

In the limit $k\to\infty$, the leading-order contribution to the retarded Fock self-energy (5.3) is

$$\Sigma^{\rm F}_{ab}(\eta,\eta',k) \sim -k^2 J(\eta,\eta')^2\, G^{\rm R}_{ab}(\eta,\eta',k)\ , \tag{5.5}$$

where

$$J(\eta,\eta')^2 = \frac{1}{6\pi^2}\int_0^\infty {\rm d}q\, \Big[P_{22}(\eta,\eta',q) + 2P_{55}(\eta,\eta',q)\Big]\ . \tag{5.6}$$

Notice that in the equal-time limit

$$J(\eta,\eta)^2 = \frac{1}{(f_+\mathcal{H})^2}\frac{C_{u_iu_i}(\eta,\eta,0)}{3}\ , \tag{5.7}$$

which is proportional to the mean-square velocity $C_{u_iu_i}(\eta,\eta,0)/3$ up to the factor $(f_+\mathcal{H})^2$, where $C_{u_iu_i}(\eta,\eta',|\boldsymbol{x}-\boldsymbol{y}|)$ is the covariance function of two velocity fields.[14] Since $J(\eta,\eta')$ depends on the full velocity power spectrum, the general solution involves solving the power spectrum equation (3.18). To avoid this complication, the full velocity power spectrum may be approximated by the linear one. In this case one obtains

$$J(\eta,\eta')^2 = {\rm e}^{\eta-\eta_{\rm in}}\,{\rm e}^{\eta'-\eta_{\rm in}}\,\sigma_{\rm v}^2\ , \tag{5.8}$$

where

$$\sigma_{\rm v}^2 = \frac{1}{6\pi^2}\int_0^\infty {\rm d}q\, P^{\rm in}_{22}(q)\ , \tag{5.9}$$

[14] Only velocity correlations enter into expression (5.6) since the dominant contribution in the limit $k\to\infty$ is the sweeping effect, which is associated with the random advection of density structures by a large-scale velocity field [35].

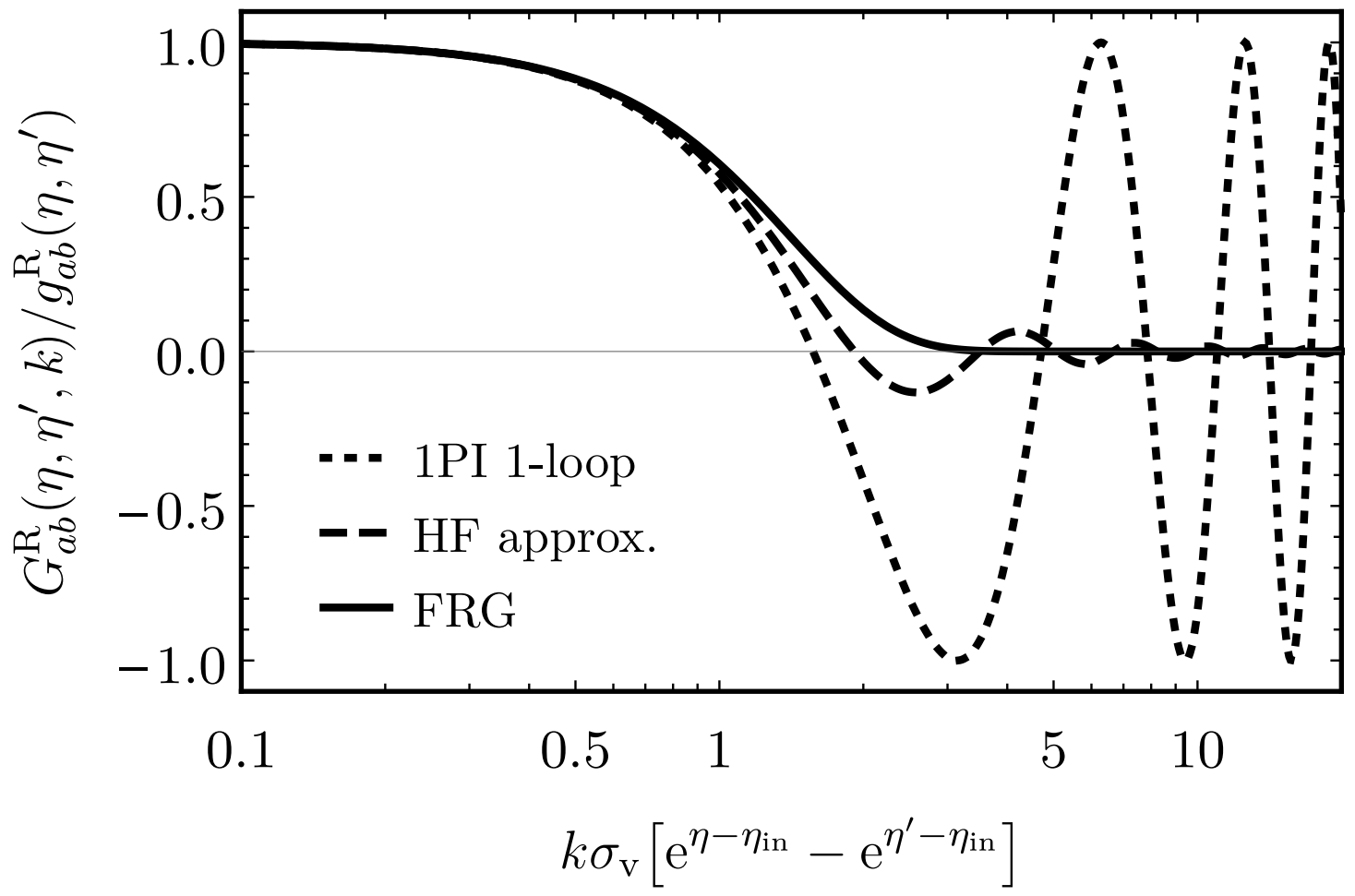

**Figure 4**. The propagators (5.12), (5.10) and (5.11) (componentwise) normalised to the linear propagator as a function of $k\sigma_{\mathrm{v}}\big[\mathrm{e}^{\eta-\eta_{\mathrm{in}}}-\mathrm{e}^{\eta'-\eta_{\mathrm{in}}}\big]$. While the functional renormalisation group (FRG) propagator (solid curve) shows a Gaussian suppression, the Hartree–Fock (HF) approximation (dashed curve) and the 1PI one-loop approximation (dotted curve) feature oscillations related to the different partial resummations of the perturbative series.

is often interpreted as a one-dimensional velocity dispersion in the sense of a mean-square velocity.

Consider the approximation where the propagator in the Fock self-energy (5.5) is evaluated at linear level. This is the 1PI one-loop approximation, which can be solved analytically [43],

$$G^{\mathrm{R}}_{ab}(\eta,\eta',k)=g^{\mathrm{R}}_{ab}(\eta,\eta')\cos\Big(k\sigma_{\mathrm{v}}\big[\mathrm{e}^{\eta-\eta_{\mathrm{in}}}-\mathrm{e}^{\eta'-\eta_{\mathrm{in}}}\big]\Big)\,, \tag{5.10}$$

where $g^{\mathrm{R}}_{ab}(\eta,\eta')$ is the linear retarded propagator. In the case where the full propagator is kept, one obtains [19, 43]

$$G^{\mathrm{R}}_{ab}(\eta,\eta',k)=g^{\mathrm{R}}_{ab}(\eta,\eta')\,\frac{J_1\big(2k\sigma_{\mathrm{v}}\big[\mathrm{e}^{\eta-\eta_{\mathrm{in}}}-\mathrm{e}^{\eta'-\eta_{\mathrm{in}}}\big]\big)}{k\sigma_{\mathrm{v}}\big[\mathrm{e}^{\eta-\eta_{\mathrm{in}}}-\mathrm{e}^{\eta'-\eta_{\mathrm{in}}}\big]}\,, \tag{5.11}$$

where $J_1$ is the first-order Bessel function of first kind.

Let us also compare this to results first obtained in the framework of renormalised perturbation theory [15, 16] and thereafter also using renormalisation group methods [31, 35]. In the large external wave number limit one obtains a resummed propagator of the form

$$G^{\mathrm{R}}_{ab}(\eta,\eta',k)=g^{\mathrm{R}}_{ab}(\eta,\eta')\,\exp\Big(-\tfrac{1}{2}\,k^2\sigma_{\mathrm{v}}^2\big[\mathrm{e}^{\eta-\eta_{\mathrm{in}}}-\mathrm{e}^{\eta'-\eta_{\mathrm{in}}}\big]^2\Big)\,, \tag{5.12}$$

with a very similar approximation using a linear instead of the full power spectrum.[15]

The large external wave number limit propagators (5.10), (5.11) and (5.12) can all be written as the linear retarded propagator multiplied by a correction factor that only depends on the dimensionless combination $k\sigma_{\mathrm{v}}\big[\mathrm{e}^{\eta-\eta_{\mathrm{in}}}-\mathrm{e}^{\eta'-\eta_{\mathrm{in}}}\big]$ and are shown in Figure 4.

The propagator obtained from the functional renormalisation group calculation (5.12) is suppressed by a Gaussian factor while the 1PI one-loop and Hartree–Fock approximation propagators feature oscillations which in the latter case are suppressed at larger $k\sigma_\mathrm{v}\big[\mathrm{e}^{\eta-\eta_\mathrm{in}} - \mathrm{e}^{\eta'-\eta_\mathrm{in}}\big]$. The characteristic scale of suppression respectively the frequency of the oscillations is determined by the *non-linear wave number*

$$k_\mathrm{nl}(\eta) = \frac{1}{\sigma_\mathrm{v}\,\mathrm{e}^{\eta-\eta_\mathrm{in}}}\ , \tag{5.13}$$

or more precisely by the difference of the inverse of two non-linear wave numbers at the two times of the propagator.

[15]Let us emphasise that this only works in the absence of velocity dispersion degrees of freedom.

The difference of the propagators is due to the different resummation schemes underlying the approximations in which they were derived. While the three methods are all non-perturbative in the sense that the perturbative series is resummed to infinite order, they correspond to different infinite partial resummations. This is best understood in the language of renormalised perturbation theory [15], where the propagator (5.12) can be obtained as a systematic resummation of the perturbative series.

The Gaussian suppression factor entering the propagator (5.12) can be written as the series

$$\mathrm{e}^{-\frac{1}{2}X^2} = \sum_{\ell=0}^{\infty} \frac{(2\ell-1)!!}{(2\ell)!}\,(-X^2)^\ell\ , \tag{5.14}$$

where $X = k\sigma_\mathrm{v}\big[\mathrm{e}^{\eta-\eta_\mathrm{in}} - \mathrm{e}^{\eta'-\eta_\mathrm{in}}\big]$ and !! is the double factorial. Here, $(2\ell-1)!!$ are the number of diagrams contributing to the propagator at perturbative loop order $\ell$ while the factor $1/(2\ell)!$ is due to all time integrations within a diagram [16]. It is emphasised that these are not all diagrams, but rather only a subclass of diagrams that are assumed to be dominant in the large external wave number limit. This is also clear from functional renormalisation group calculations, where the propagator (5.12) was obtained in an approximation [35]. As such, it corresponds to an infinite partial resummation of the perturbative series.

Applying the same expansion to the propagators (5.10) and (5.11), one can find the number of diagrams contributing at each loop order. For the 1PI one-loop approximation the expansion of the cosine reads

$$\cos(X) = \sum_{\ell=0}^{\infty} \frac{1}{(2\ell)!}\,(-X^2)^\ell\ , \tag{5.15}$$

while for the Hartree–Fock approximation one finds

$$\frac{J_1(2X)}{X} = \sum_{\ell=0}^{\infty} \left[\frac{(2\ell)!}{(\ell+1)!\,\ell!}\right] \frac{1}{(2\ell)!}\,(-X^2)^\ell\ . \tag{5.16}$$

The number of diagrams contributing at each loop order to the different resummation schemes is summarised in Table 1. It is noteworthy that in the direct-interaction approximation similar results for the propagator hold in turbulence [50, 51].

The different partial resummations can also be obtained from topological arguments. Consider the contributions to the propagator contained in the resummation of renormalised perturbation theory up to two-loop order. At zero- and one-loop order there is only a single diagram which is captured by all resummation schemes. At two-loop order though there are the three diagrams

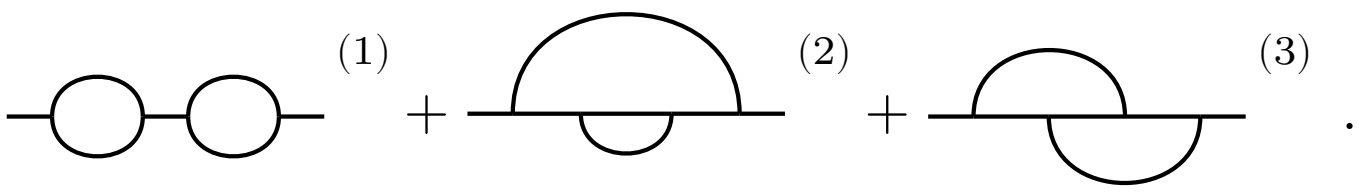

(5.17)

| Loop order | 1 | 2 | 3 | 4 | 5 | ... | $\ell$ |
|---|---|---|---|---|---|---|---|
| 1PI 1-loop | 1 | 1 | 1 | 1 | 1 | ... | 1 |
| HF approx. | 1 | 2 | 5 | 14 | 42 | ... | $(2\ell)!/((\ell+1)!\ell!)$ |
| FRG | 1 | 3 | 15 | 105 | 945 | ... | $(2\ell-1)!!$ |

**Table 1**. Number of diagrams included at loop order $\ell$ in the different resummation schemes. While the 1PI one-loop approximation captures only a single diagram at each loop order, the Hartree–Fock approximation improves on this, but captures only a subset those diagrams resummed in the functional renormalisation group (FRG).

While the 1PI one-loop approximation can only capture diagram (1) topologically, the Hartree–Fock approximation contains the diagrams (1) and (2) due to its self-consistent one-loop structure. Finally, the functional renormalisation group calculation captures all three diagrams (1), (2) and (3) [35].

Based on the large external wave number limit, two approximation schemes interpolating between the perturbative small wave number and non-perturbative large wave number sector have been developed. In renormalised perturbation theory the propagator (5.12) is used for the interpolation [16], while in the direct-interaction approximation the propagator (5.11) is utilised [19]. In the next section the numerical solution of the full propagator equation (3.17) is compared to these interpolation schemes, which in the following are referred to as 'CS' and 'TH' approximations, respectively.

### 5.2 Numerical solutions

In this section the Hartree–Fock approximation is solved with numerical methods and compared to $N$-body simulations. In the following, it is convenient to work with dimensionless power spectra,

$$\Delta_{ab}(\eta, \eta', k) = \frac{k^3}{2\pi^2} P_{ab}(\eta, \eta', k) \ . \tag{5.18}$$

#### 5.2.1 $N$-body simulations

Since $N$-body simulations can only simulate a representative finite region of the Universe, one typically has to decide whether one wants to accurately reproduce large-scale structures, requiring a fairly large simulation volume, or is interested in resolving small-scale physics, which requires a rather high number density of particles. To test the performance of the Hartree–Fock approximation, numerical solutions are compared to the Horizon Run 2 (HR2) $N$-body simulation [56], featuring a rather large simulation volume, as well as to a $N$-body simulation of Buehlmann & Hahn (BH19) [59], featuring a comparably high number density of particles. The cosmological parameters and codes used for the simulations as well as the number of particles $N_\mathrm{p}$, the box side length $L_\mathrm{box}$ and the initial redshift $z_\mathrm{in}$ are summarised in Table 2.

Three snapshots of the BH19 $N$-body simulation were kindly provided by the authors M. Buehlmann and O. Hahn [59]. The datasets contain estimates for the density, velocity and isotropic velocity dispersion, interpolated on a $1024^3$-mesh at redshifts $z = 2.165$, $z = 0.994$ and $z = 0$. The density field was estimated using the cloud-in-cell deposition algorithm

| Simulation | Cosmology | Code | $N_\mathrm{p}$ | $L_\mathrm{box}$ [Mpc/$h$] | $z_\mathrm{in}$ |
|---|---|---|---|---|---|
| HR2 [56] | WMAP5 [57] | `gotpm` [58] | $6000^3$ | 7200 | 32 |
| BH19 [59] | Planck 2015 [60] | `gadget-2` [61] | $1024^3$ | 300 | 99 |

**Table 2**. Details of the $N$-body simulations used as a comparison to the numerical solutions of the Hartree–Fock approximation.

[63], while the velocity and velocity dispersion fields were obtained from the Lagrangian tessellation method [64, 65].

To derive an estimate for the power spectra, the fields are discretely Fourier transformed on a three-dimensional grid with fundamental wave number $k_\mathrm{f} = 2\pi/L_\mathrm{box}$ and $N_\mathrm{grid} = 1024$ grid points per dimension. Further, the density field is deconvolved using the cloud-in-cell window function [66]

$$W(\boldsymbol{n}) = \prod_{i=1}^{3} \mathrm{sinc}\left(\frac{\pi n_i}{N_\mathrm{grid}}\right)^2 , \tag{5.19}$$

where $n_i \in \{1, \dots, N_\mathrm{grid}\}$. The power spectrum estimate $\hat{P}_{ab}(k)$ is obtained from the spatial average of two fields over the simulation volume and an angular average of modes within spherical shells of thickness $k_\mathrm{f}$ and radius $k = nk_\mathrm{f}$ for $n \in \{1, \dots, N_\mathrm{grid}\}$. Finally, the statistical error is estimated as [67]

$$\Delta\hat{P}_{ab}(k) = \frac{1}{\sqrt{2\pi n}}\left[\hat{P}_{ab}(k) + \left(\frac{L_\mathrm{box}}{N_\mathrm{grid}}\right)^3\right] , \tag{5.20}$$

where the first term on the right-hand side is the sample variance, while the second term is the Poisson shot noise. It is emphasised that the power spectrum and its error estimate should be taken with caution for small wave numbers due to systematic uncertainties related to the relatively small box size of the BH19 $N$-body simulation.

Similarly, an estimate of the velocity dispersion mean field can be computed from the spatial average of the trace of the velocity dispersion tensor.

### 5.2.2 Numerical methods

The system of equations (3.16), (3.17) and (3.18) in the Hartree–Fock approximation (5.3) and (5.4) is numerically solved in time using an adaptive Runge–Kutta–Cash–Karp method [52–54] for Volterra integro-differential equations [55] and in wave number using a finite element method with B-splines of order unity as basis functions [20, 43]. To test the Hartree–Fock approximation against the HR2 and BH19 $N$-body simulations, two sets of numerical solutions are computed. These account for the different cosmological parameters and initial redshifts listed in Table 2.

The initial power spectrum is generated on scales $10^{-5}\ h/\mathrm{Mpc} \lesssim k \lesssim 20\ h/\mathrm{Mpc}$ using the `class` code [62] and extrapolated into the ultraviolet with the fitting formula $\alpha \log(\beta k)^2 k^{n_\mathrm{s}-4}$, where $n_\mathrm{s}$ is the scalar spectral index. To ensure comparability to the HR2 and BR19 $N$-body simulations, wave numbers are interpolated on a grid with an infrared cut-off $k_\mathrm{min} = k_\mathrm{f}$ and an ultraviolet cut-off $k_\mathrm{max} = nk_\mathrm{Ny}$ equal to a multiple of the Nyquist wave number $k_\mathrm{Ny} = \pi N_\mathrm{grid}/L_\mathrm{box}$. For the HR2 simulation these are given by $k_\mathrm{f} \approx 8.7 \cdot 10^{-4}\ h/\mathrm{Mpc}$ and $k_\mathrm{Ny} \approx 2.6\ h/\mathrm{Mpc}$, while for the BH19 simulation these are $k_\mathrm{f} \approx 2.1 \cdot 10^{-2}\ h/\mathrm{Mpc}$ and $k_{\rm Ny} \approx 11\ h/\text{Mpc}$. More specifically, when comparing numerical solutions to the HR2 simulation, the cut-off is chosen as $k_{\rm max} \approx 5.2\ h/\text{Mpc}$, while when comparing to the BH19 simulation the cut-off is varied between $k_{\rm max} \approx 11\ h/\text{Mpc}$ and $k_{\rm max} \approx 43\ h/\text{Mpc}$.[16]

For the single-stream approximation a wave number grid with $N_k = 100$ interpolation points is used, while for the full set of fields (3.1) a grid with $N_k = 50$ is chosen. While it would be desirable to increase the number of interpolation points, it considerably slows down the solving algorithm since the arrays storing the Fock self-energies (5.3) and (5.4) scale cubic in $N_k$.[17]

### 5.2.3 Single-stream approximation

To gain a better understanding of the properties of the Hartree–Fock approximation, the equations (3.17) and (3.18) are solved in the absence of a mean field in the single-stream approximation, where the field content (3.1) reduces to the density and velocity-divergence only.

The density contrast propagator and power spectrum have also been studied using the two-particle irreducible method [42, 43, 48, 68] and closure theory [19–21], which yield the same evolution equations as the Hartree–Fock approximation in the single-stream approximation. The purpose of the following is to identify how well the Hartree–Fock approximation captures non-linearities and to find out where the single-stream approximation fails.

**Propagator.** In Figure 5, the time (upper panel) and wave number (lower panel) dependencies of the density contrast and velocity-divergence reduced propagators are shown at constant wave number $k \approx 2\ h/\text{Mpc}$ and at fixed redshift $z = 0$, respectively. The numerical solution of the Hartree–Fock approximation is near to identical with the TH approximation. A slight deviation for wave numbers $k \gtrsim k_{\rm nl}(\eta)$ is observed, which is not too surprising since the TH approximation interpolates between the small and large wave number sector. As was already discussed in Section 5.1, the observed oscillations are not associated with real physical phenomena but are rather due to the partial resummation of the full perturbative series.

**Density contrast equal-time auto-spectrum.** In Figure 6, the dimensionless density contrast equal-time auto-spectrum, normalised to the linear spectrum, is shown at redshifts $z = 2$ (upper panel) and $z = 0$ (lower panel). At large wave numbers, the Hartree–Fock approximation does not overestimate the $N$-body power spectrum as bad as the standard perturbation theory one-loop prediction. While the deviation from the $N$-body power spectrum is fairly small at redshift $z = 2$, it grows towards redshift $z = 0$. Especially for wave numbers $k \lesssim 0.2\ h/\text{Mpc}$ the Hartree–Fock approximation seems to follow the standard perturbation theory one-loop prediction. Since the full 1PI three-point functions are set to their bare form in the Hartree–Fock approximation, it is likely that vertex corrections are needed to accurately capture the mildly non-linear regime.

In Figure 7, the dimensionless density contrast equal-time auto-spectrum is shown for a larger range of wave numbers at redshift $z = 0$. The Hartree–Fock approximation shows a much better convergence towards the $N$-body power spectrum compared to the standard perturbation theory one-loop prediction, which badly overestimates the power spectrum. The drop of the $N$-body power spectrum at wave numbers near the Nyquist wave number is due to finite box size effects and should therefore not be trusted.

[16] While numerical solutions computed in the single-stream approximation are percent-accurate for $k_{\rm max} \gtrsim 5\ h/\text{Mpc}$ [19], this changes dramatically when including velocity dispersion degrees of freedom, since the mean field source term (3.22) is sensitive to small-scale physics.

[17] The self-energy arrays also scale with the sixth power in the field content, which is the reason why the vector and tensor velocity dispersion modes are neglected.

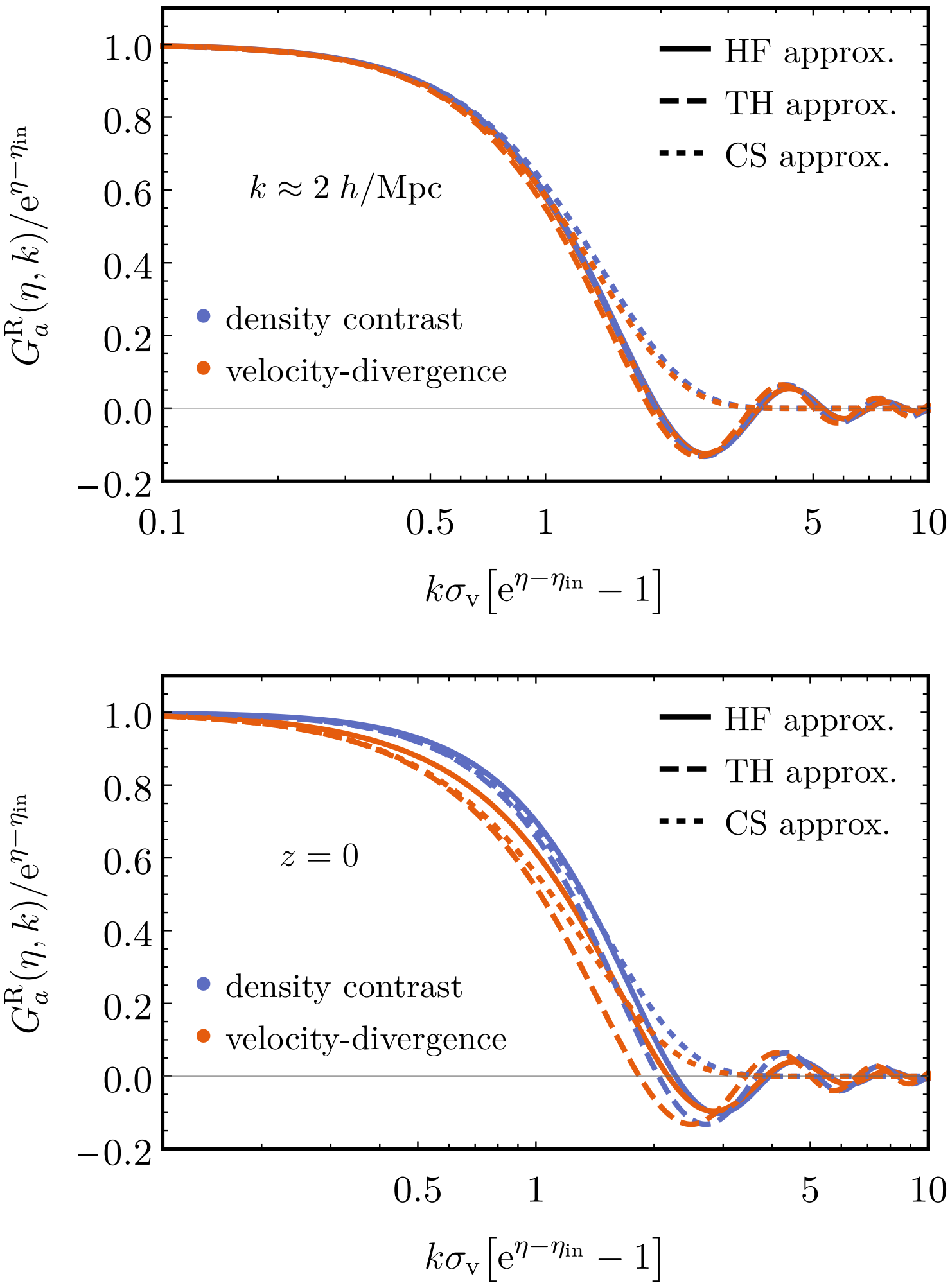

**Figure 5**. Single-stream approximation reduced density contrast (blue curves) and velocity-divergence (red curves) propagator time dependence at $k \approx 2\ h/\mathrm{Mpc}$ (upper panel) and wave number dependence at $z = 0$ (lower panel). The numerical solution of the Hartree–Fock (HF) approximation (solid curves) is compared to the TH (dashed curves) and CS (dotted curves) approximation. The Hartree–Fock and TH approximations are near to identical since the latter is based on the large wave number limit of the former.

In Figure 8, the dimensionless density contrast equal-time auto-spectrum (upper panel) is shown at redshift $z = 0$. However, here it is computed with different cosmological parameters and cut-offs, as it is being compared to the BH19 $N$-body simulation. Here, the standard perturbation theory one-loop prediction is actually performing better than the Hartree–Fock approximation, which is due to the comparably large infrared cut-off $k_{\mathrm{min}} \approx 2.1 \cdot 10^{-2}\ h/\mathrm{Mpc}$ that is resulting in smaller non-linear corrections to the power spectrum compared to the overestimation observed in Figure 7, where the infrared cut-off is much smaller.

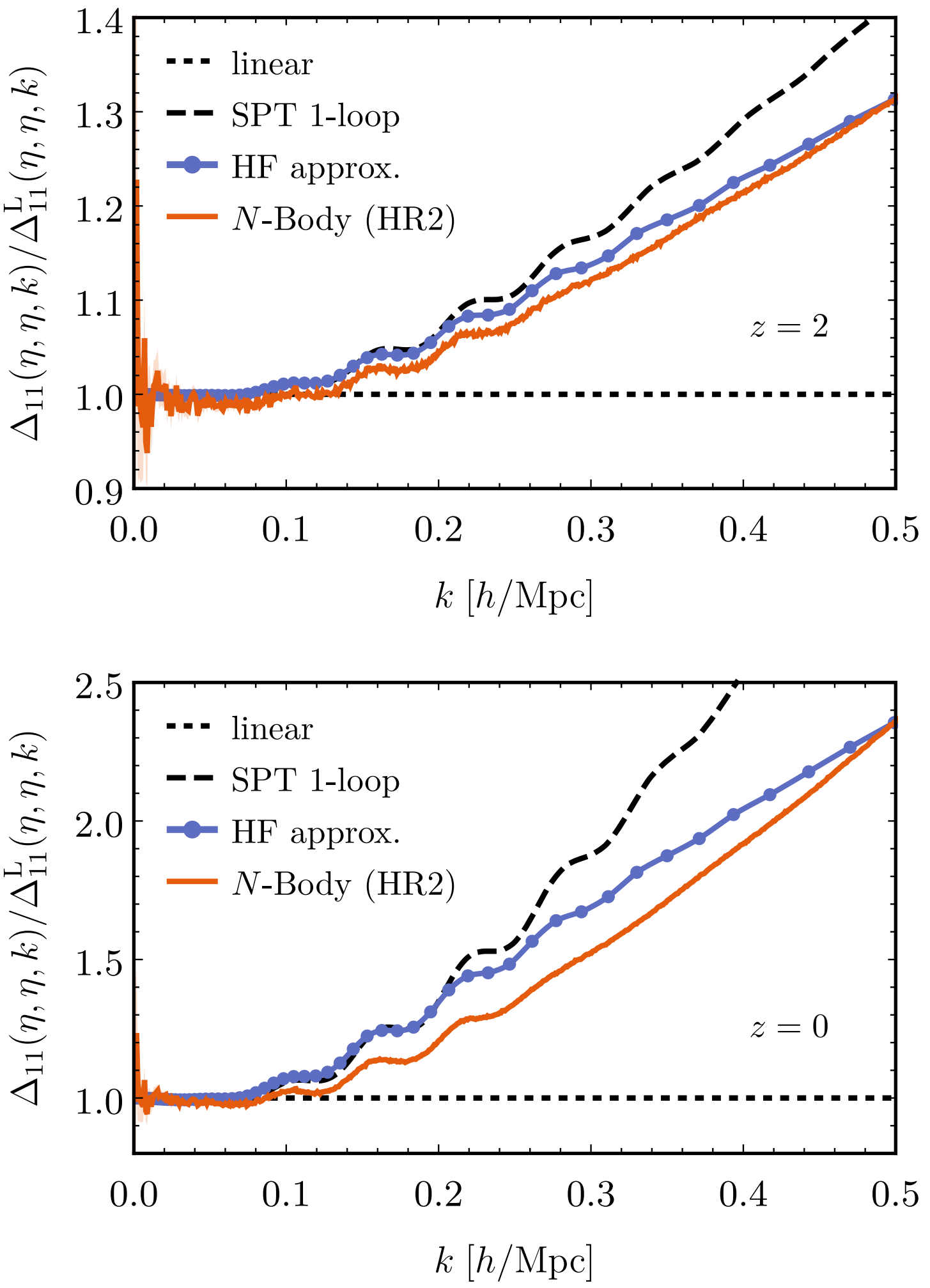

**Figure 6**. Single-stream approximation dimensionless density contrast equal-time auto-spectrum, normalised to the linear spectrum (black dotted curves), at redshifts $z = 2$ (upper panel) and $z = 0$ (lower panel). The numerical solution of the Hartree–Fock (HF) approximation (blue solid curves) is compared to the standard perturbation theory (SPT) one-loop prediction (black dashed curves) and data from the HR2 $N$-body simulation (red solid curves). While the Hartree–Fock approximation shows better convergence properties at larger wave numbers, it overestimates the power spectrum at smaller wave numbers.

**Velocity-divergence equal-time auto-spectrum.** In Figure 8, the velocity-divergence equal-time auto-spectrum (lower panel) is shown at redshift $z = 0$. Neither the Hartree–Fock nor the standard perturbation theory one-loop prediction can capture the sharp drop of the velocity-divergence power spectrum at small scales. Even worse, the Hartree–Fock approximation even fails to reproduce the power spectrum at relatively small wave numbers, where the standard perturbation theory one-loop prediction can capture at least the onset of the suppression. Indeed, the drop is associated with the energy transfer into vorticity modes [69] and is thus beyond the single-stream approximation. This shortcoming can be resolved by including vorticity and velocity dispersion degrees of freedom and is discussed in greater detail in Section 5.2.4.

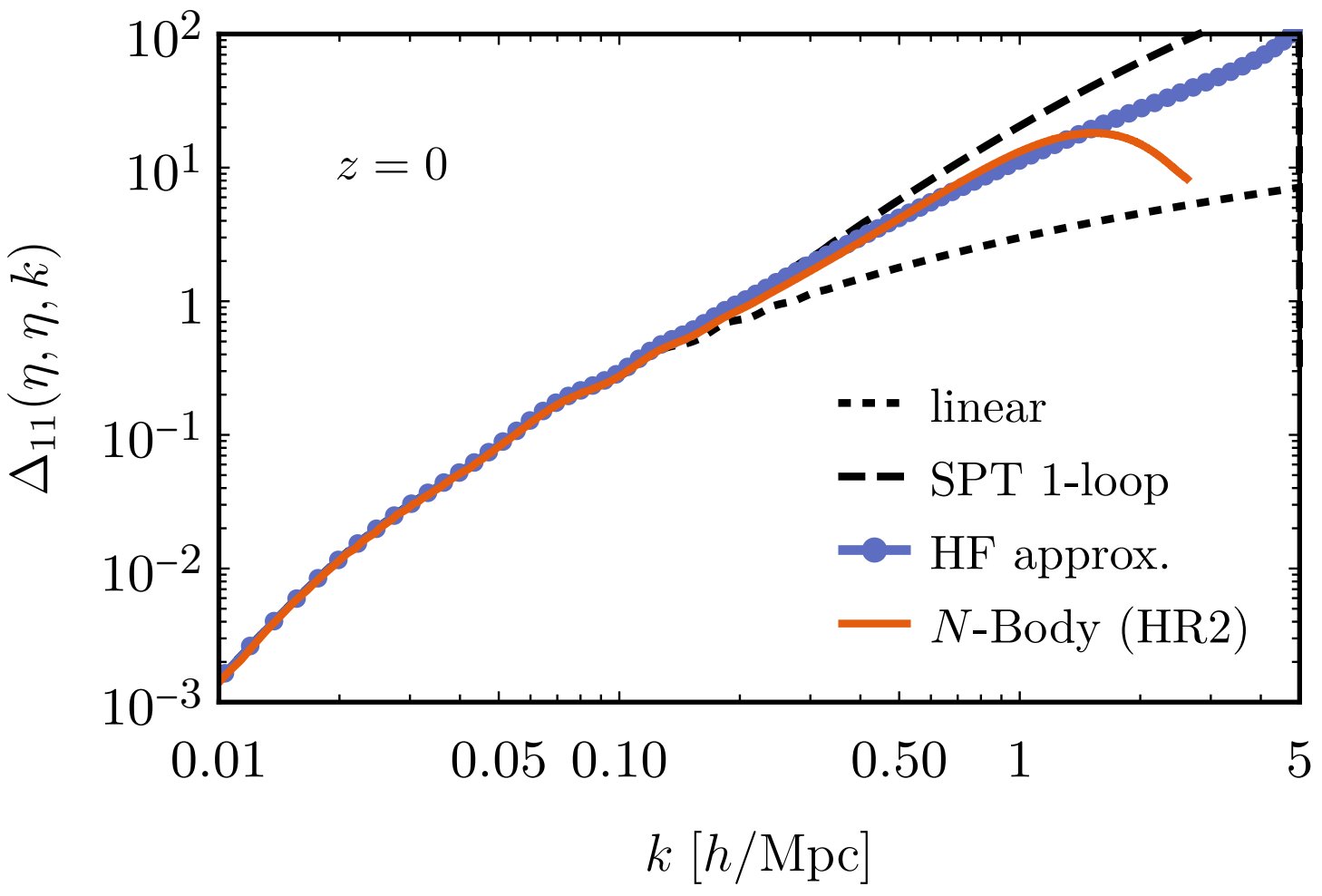

**Figure 7**. Single-stream approximation dimensionless density contrast equal-time auto-spectrum at redshift $z = 0$. The Hartree–Fock approximation captures the general shape of the non-linear power spectrum obtained from $N$-body simulations much better than the standard perturbation theory one-loop prediction.

**Unequal-time auto-spectra and equal-time cross-spectra.** In Figure 9, the dimensionless density contrast unequal-time auto-spectrum (upper panel) at redshifts $z = 0$ and $z' = 2.165$, as well as the dimensionless velocity-divergence-density contrast equal-time cross-spectrum (lower panel) at redshift $z = 0$, are shown. The oscillatory suppression of the unequal-time power spectrum is qualitatively captured by the Hartree–Fock approximation, although this is likely due to the partial resummation scheme that is also responsible for the unphysical oscillations in the propagator seen in Figure 5.

More interesting and important for the investigations in the next section that include velocity dispersion degrees of freedom is the cross-spectrum. The $N$-body power spectrum stays close to the linear prediction up to $k \approx 1.0\, h/\text{Mpc}$ above which a sign change is observed. This is interpreted as a turnover from matter inflow to outflow and associated with shell-crossing [59, 70]. Naturally, this cannot be captured in the single-stream approximation and explains why neither the standard perturbation theory result nor the Hartree–Fock approximation are able to capture the $N$-body power spectrum.

**Summary.** In summary, the insights gained within the single-stream approximation are:

- The propagators are to very good approximation described by the large wave number limit (5.11), at least if the initial and final times are sufficiently far apart.
- The density contrast equal-time auto-spectrum converges well at small scales but fails to accurately capture the physics at mildly non-linear scales. This is most likely due to neglecting corrections to the bare vertex.
- The drop of the velocity-divergence equal-time auto-spectrum cannot be captured since it is associated to the energy transfer into vorticity modes.
- The oscillatory suppression of unequal-time auto-spectra is qualitatively captured due to the non-perturbative resummation of the propagators.

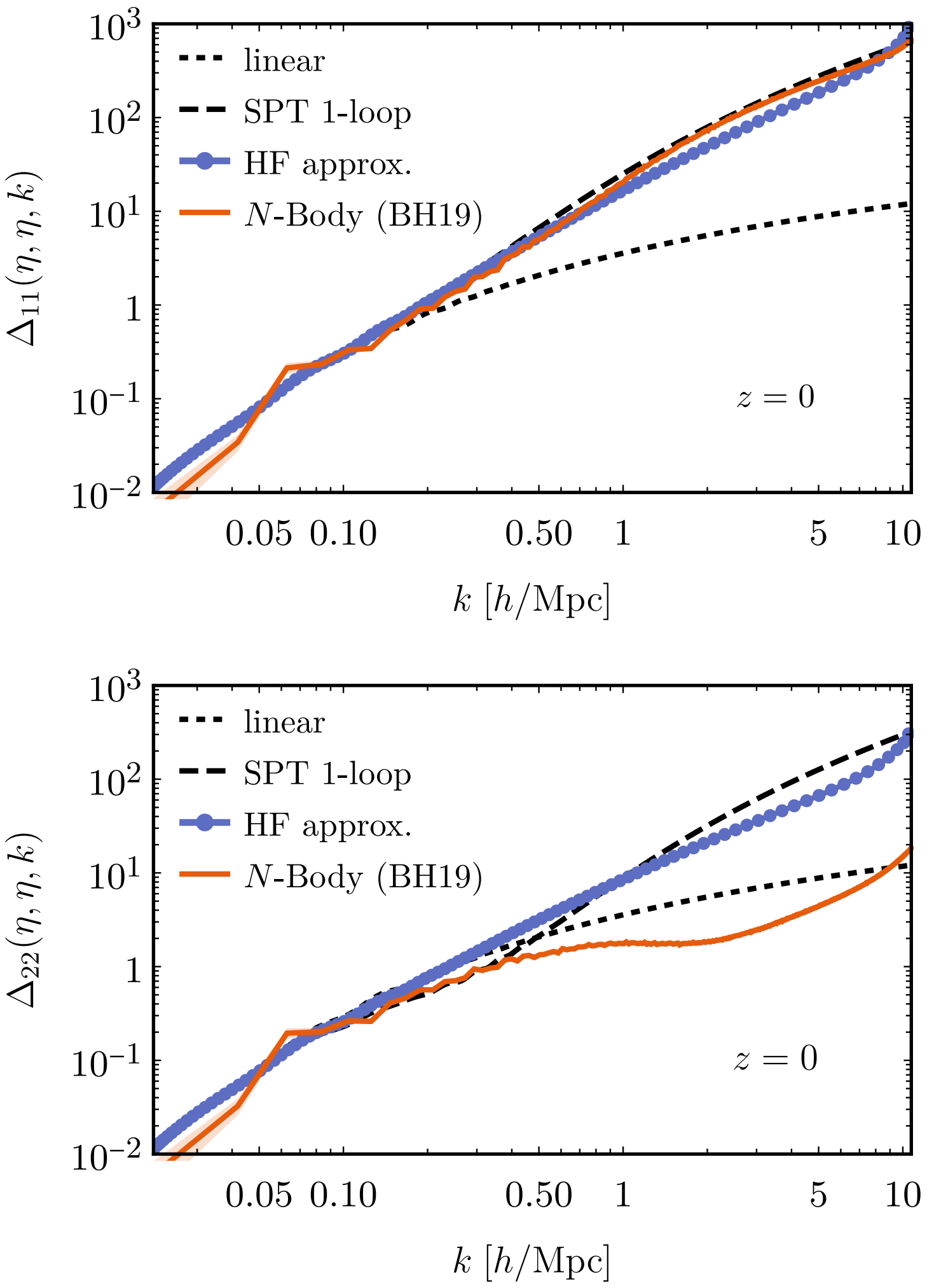

**Figure 8**. Single-stream approximation dimensionless density contrast (upper panel) and velocity-divergence (lower panel) equal-time auto-spectrum at redshift $z = 0$. The numerical solution of the Hartree–Fock (HF) approximation (blue solid curves) is compared to the standard perturbation theory (SPT) one-loop prediction (black dashed curves) and data from the BH19 $N$-body simulation (red solid curve). The Hartree–Fock approximation badly overestimates the $N$-body velocity-divergence power spectrum, even in the small wave number regime. The drop of the velocity-divergence power spectrum is associated with the energy transfer into vorticity modes and is thus beyond single-stream approximation.

- The sign change in the velocity-divergence-density contrast equal-time cross-spectrum cannot be captured since it is associated with shell-crossing.

#### 5.2.4 Including velocity dispersion degrees of freedom

Having seen the limitations of the single-stream approximation in the last section, the Hartree–Fock approximation is now studied for the field content (3.1), including the vorticity and velocity dispersion fields. In this case, the free-streaming of dark matter seen in the Hartree approximation as well as the non-linear coupling of modes encoded in the Fock self-energies contribute. To take both effects correctly into account, the two scales associated with them, namely the free-streaming wave number $k_{\text{fs}}$ and the non-linear wave number $k_{\text{nl}}$, need to be resolved. Numerically this is rather challenging for cold dark matter candidates because these scales are separated by several orders of magnitudes. Because the propagators are suppressed on small scales, the adaptive Runge–Kutta method takes smaller times steps as one solves at later times. The larger the ultraviolet cut-off wave number is, the more time steps are necessary to keep the errors under control. To obtain results in a sensible amount of time, but at the same time capture the effects of non-vanishing velocity dispersion, rather warm dark matter candidates are studied in the following. More specifically the initial velocity dispersion mean field is taken to be between $\bar{\sigma}^{\text{in}} = 10^2\ \text{km}^2/\text{s}^2$ and $\bar{\sigma}^{\text{in}} = 10^5\ \text{km}^2/\text{s}^2$.

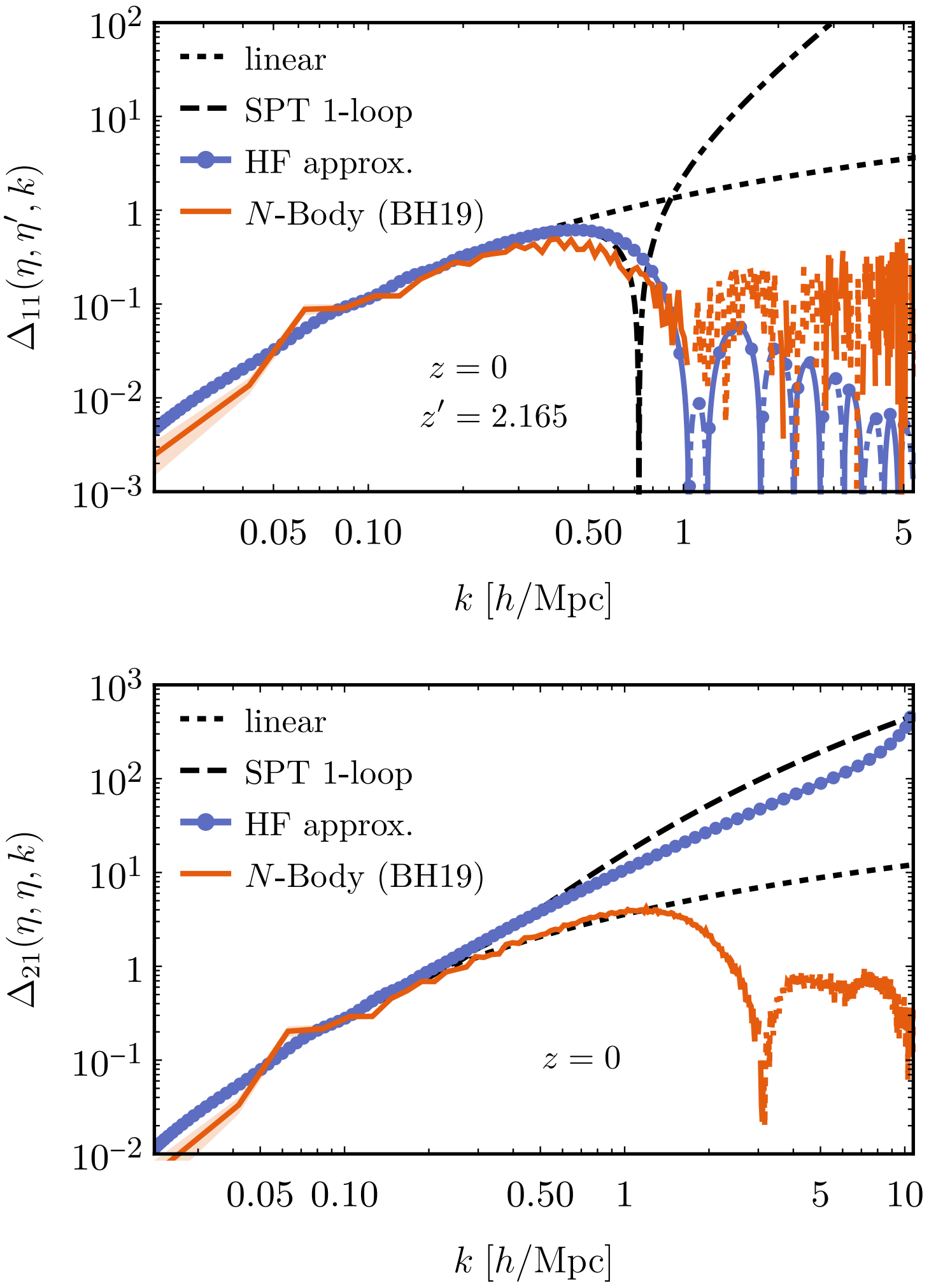

**Figure 9**. Single-stream approximation dimensionless density contrast unequal-time auto-spectrum (upper panel) at redshifts $z = 0$ and $z' = 2.165$ as well as velocity-divergence-density contrast equal-time cross-power spectrum (lower panel) at redshift $z = 0$. The numerical solution of the Hartree–Fock (HF) approximation (blue solid curves) is compared to the standard perturbation theory (SPT) one-loop prediction (black dashed curves) and data from the BH19 $N$-body simulation (red solid curve). The oscillatory suppression of the unequal-time power spectrum is qualitatively captured by the Hartree–Fock approximation, while the sign change observed in the equal-time cross-spectrum cannot be captured since it is associated with shell-crossing.

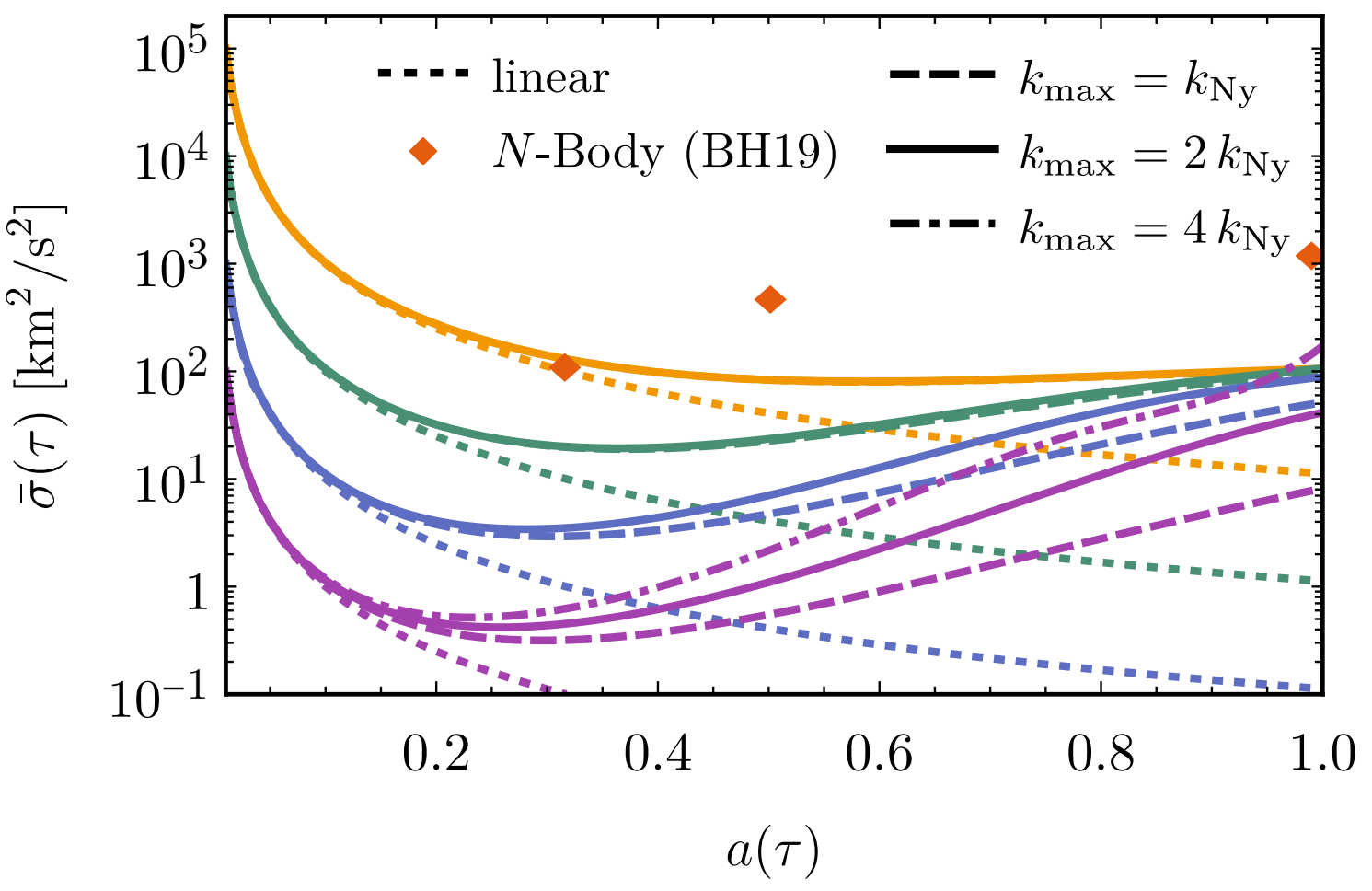

**Figure 10**. Evolution of the velocity dispersion mean field in the Hartree–Fock approximation for different ultraviolet cut-offs that are multiples of the Nyquist wave number $k_{\rm Ny} \approx 11\ h/{\rm Mpc}$. Shown are colour-coded numerical solutions for differential initial conditions (violet, blue, green and yellow from coldest to warmest), the linearly decaying mean field (dotted curves) and three snapshots from the BH19 $N$-body simulation (red diamonds). For warmer dark matter, where the free-streaming wave number is below the cut-off, the results are well converged, while for colder dark matter with a larger free-streaming wave number this is no longer the case.

The ultraviolet cut-off is varied between $k_{\rm max} = k_{\rm Ny}$ and $k_{\rm max} = 4k_{\rm Ny}$ to study the effect of including smaller scales.

**Velocity dispersion mean field.** In Figure 10, the velocity dispersion mean field for different ultraviolet cut-offs is shown. The different initial conditions are colour-coded in a scheme that is also used in all following power spectrum figures. The numerical solutions have been obtained for the ultraviolet cut-offs $k_{\rm max} = k_{\rm Ny}$ and $k_{\rm max} = 2k_{\rm Ny}$ as well as for $k_{\rm max} = 4k_{\rm Ny}$ for the coldest candidate with an initial velocity dispersion mean field of $\bar{\sigma}^{\rm in} = 10^2\ {\rm km}^2/{\rm s}^2$.

As long as the free-streaming wave number is below or of the order of the ultraviolet cut-off, the numerical solutions are well converged and insensitive to a larger ultraviolet cut-off. This is best seen for the warmest candidate with an initial velocity dispersion mean field of $\bar{\sigma}^{\rm in} = 10^5\ {\rm km}^2/{\rm s}^2$. The corresponding free-streaming wave number is $k_{\rm fs} \approx 2.1\ h/{\rm Mpc}$ at $z = 99$ and grows to $k_{\rm fs} \approx 6.2\ h/{\rm Mpc}$ at $z = 0$, well below the ultraviolet cut-off. For colder initial conditions, one naturally needs a larger ultraviolet cut-off to capture the small-scale power that sources the growth of the mean field. This can be seen best for the coldest candidate, where the free-streaming wave number at the point of turnover from decay to growth is roughly of the order $k_{\rm fs} \approx 210\ h/{\rm Mpc}$, which is way above the largest ultraviolet cut-off $k_{\rm max} = 4k_{\rm Ny} \approx 43\ h/{\rm Mpc}$ used here. Naturally, this extends even more drastically to colder candidates, which increases the need to include ever larger wave numbers.

Comparing to the BH19 $N$-body simulation, it is evident that the velocity dispersion mean field is even larger than the warmest dark matter candidates studied here. The similar velocity dispersion mean field values at late times for dark matter candidates with very different initial conditions suggest some universal fixed point at late times. The inability to quantitatively predict the simulation data is likely related to the Hartree–Fock approximation or the restricted set of degrees of freedom. On the other hand, testing this with very cold dark matter candidates, such as a weakly interacting particle with $\bar{\sigma}^{\mathrm{in}} \sim 10^{-10}$ km$^2$/s$^2$, would be necessary in order to rule out the indicated very strong growth behaviour that is observed in the Hartree approximation in Figure 2, which suggests that candidates that start out colder might become warmer at later times than candidates that initially start out warmer. Due to the above mentioned computational limitations it was not possible to verify this here.

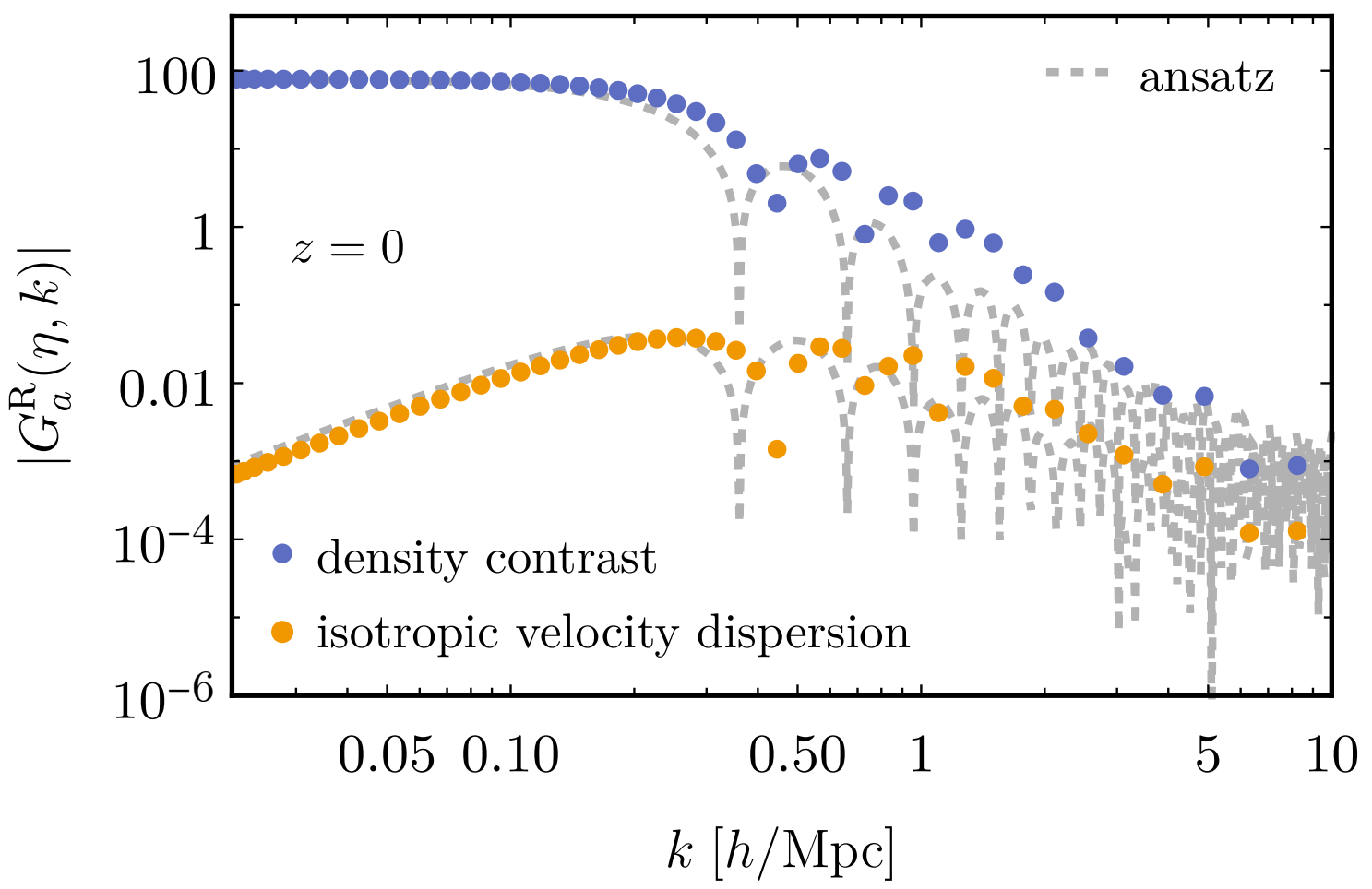

**Figure 11**. Density contrast and isotropic velocity dispersion reduced propagators for a dark matter model with $\bar{\sigma}^{\mathrm{in}} = 10^5$ km$^2$/s$^2$ at redshift $z = 0$. Also shown is the ansatz (5.21), superposing the effect of free-streaming found in the Hartree approximation (4.8) and the ultraviolet suppression due to the sweeping effect in the large wave number limit (5.11). Qualitatively, the ansatz captures the numerical solution of the Hartree–Fock approximation which naturally captures both effects.

**Propagator.** In Figure 11, the density contrast and isotropic velocity dispersion reduced propagators for the dark matter model with $\bar{\sigma}^{\mathrm{in}} = 10^5$ km$^2$/s$^2$ are shown at redshift $z = 0$. To qualitatively understand the wave number dependence of the propagator, the ansatz

$$G_{ab}^{\mathrm{R}}(\eta, \eta', k) = \tilde{G}_{ab}^{\mathrm{R}}(\eta - \eta', k^2\tilde{\sigma})\, \frac{J_1\big(2k\sigma_{\mathrm{v}}\big[\mathrm{e}^{\eta-\eta_{\mathrm{in}}} - \mathrm{e}^{\eta'-\eta_{\mathrm{in}}}\big]\big)}{k\sigma_{\mathrm{v}}\big[\mathrm{e}^{\eta-\eta_{\mathrm{in}}} - \mathrm{e}^{\eta'-\eta_{\mathrm{in}}}\big]}\,, \tag{5.21}$$

is also shown. It is based on the large external wave number limit (5.11), where $\tilde{G}_{ab}^{\mathrm{R}}(\eta - \eta', k^2\tilde{\sigma})$ is the retarded propagator that is obtained in the Hartree approximation for a linearly decaying mean field and can be reconstructed from the three independent solutions (4.8). This naïve ansatz qualitatively captures the Hartree–Fock approximation and reproduces the superposition of oscillations from free-streaming encoded in the Hartree self-energy (3.24) and the decaying oscillations due to the coupling of modes encoded in the retarded Fock self-energy (5.3).[18] Although not shown here, the performance of the ansatz (5.21) worsens for colder dark matter models. Since the propagator $\tilde{G}_{ab}^{\mathrm{R}}(\eta - \eta', k^2\tilde{\sigma})$ is strictly speaking only valid for a linearly decaying mean field, it fails to capture the effect of a strongly growing mean field, as is the case for colder dark matter models.

[18] It is emphasised that the oscillations due to the retarded Fock self-energy are a consequence of the employed truncation of the Dyson–Schwinger hierarchy and can be understood as an infinite but partial resummation of standard perturbation theory.

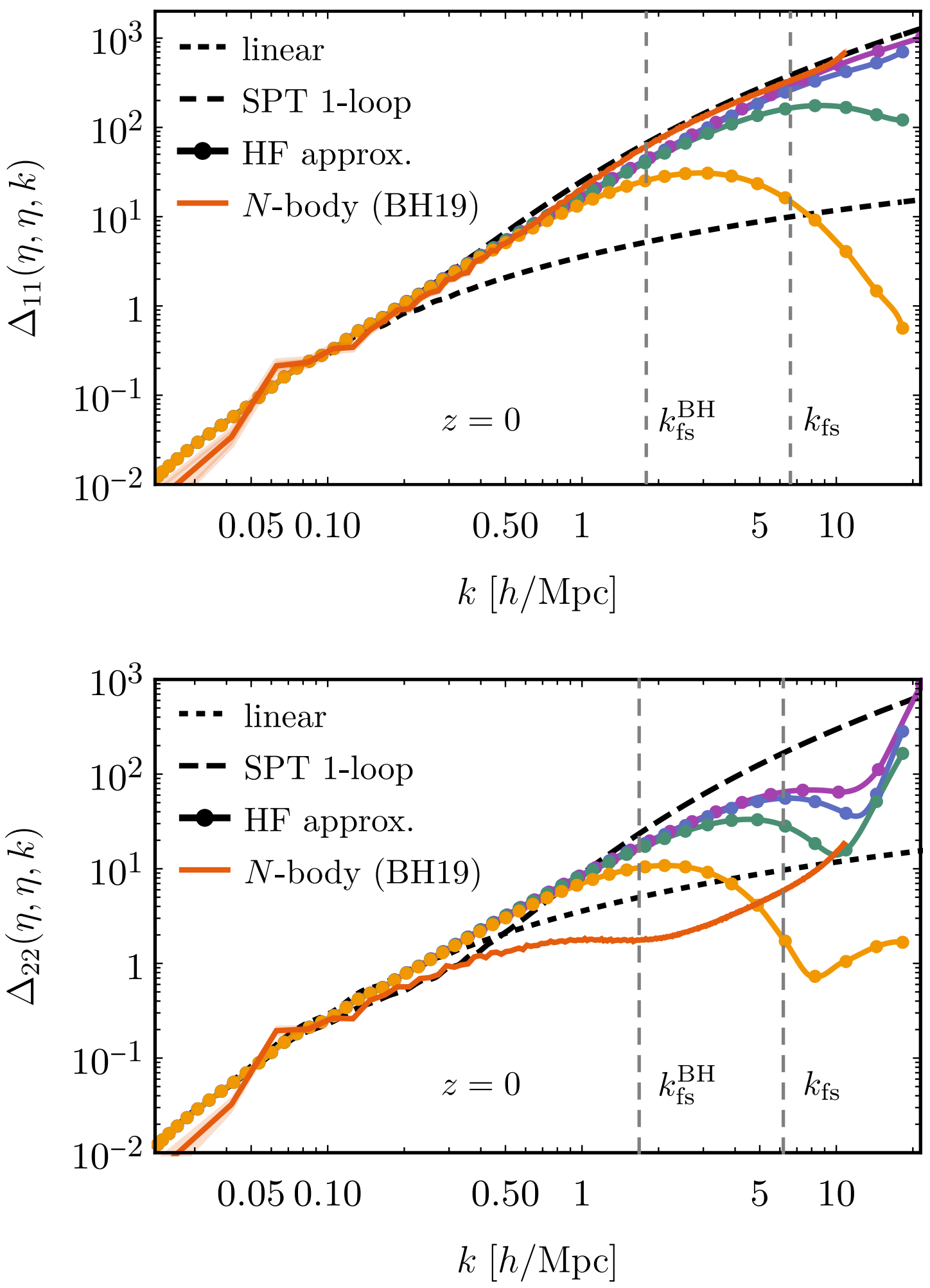

**Figure 12**. Dimensionless density contrast (upper panel) and velocity-divergence (lower panel) equal-time auto-spectrum at redshift $z = 0$. The numerical solutions of the Hartree–Fock (HF) approximation are colour coded according to the respective mean fields shown in Figure 10. They are compared to the linear power spectrum (black dotted curves), the standard perturbation theory (SPT) one-loop prediction (black dashed curves) and data from the BH19 $N$-body simulation (red solid curve). Both spectra are suppressed for warmer dark matter models, although the velocity-divergence power spectrum shows a much stronger suppression.

**Density contrast and velocity-divergence equal-time auto-spectra.** Although only initially rather warm dark matter candidates are studied, the question arises whether the approximation including velocity dispersion degrees of freedom can overcome the shortcomings of the single-stream approximation. To see the impact on non-vanishing velocity dispersion, the dimensionless density contrast and velocity-divergence equal-time auto-spectra are shown in Figure 12. The solutions correspond to different velocity dispersion mean field initial conditions and are colour-coded in the same way as in Figure 10. Additionally, the free-streaming scales related to the corresponding velocity dispersion mean field are displayed. Since the numerical solutions all have a mean field of the same order at redshift $z = 0$, and therefore a similar free-streaming wave number, only a single free-streaming wave number is indicated here. It is clearly visible that the dark matter models starting out with a warmer initial condition show a large suppression in the density contrast as well as in the velocity-divergence power spectrum. Interestingly, one can observe that although the velocity dispersion is of similar order at redshift $z = 0$, the dark matter models starting out comparably cold show near to no suppression in the density contrast power spectrum and match the $N$-body data rather well. Comparing with the evolution of the mean field shown in Figure 10, it seems that if the mean field obtains a large value only at late times, one does not see a suppression in the density contrast power spectrum. In contrast, consider the warmest dark matter model shown (orange lines) that is warm for a large part of its time evolution. Although the final mean field is even below the final mean field of the initially colder model, a significant suppression is observed in the density contrast power spectrum. This suggest that indeed the dark matter model should start out with a small velocity dispersion mean field in order to match the observed power spectrum. On the other hand, this also implies that a late-time large velocity dispersion mean field does not necessarily imply that the density contrast power spectrum needs to be suppressed.

This is rather different for the velocity-divergence power spectrum, which shows a dip in the equal-time power spectrum for all dark matter models. Although the power spectrum does not match the $N$-body simulations quantitatively, the qualitative behaviour of the dip can be understood from the numerical solutions obtained here. As the suppression of the velocity-divergence power spectrum in the mildly non-linear regime is related to the energy transfer into vorticity modes [69], the position of the suppression dip is sensitive to the vorticity power spectrum. As is shown in Figure 15 and argued below, the amplitude of vorticity spectrum is heavily dependent on the magnitude of the velocity dispersion mean field, which is underestimated in the numerical solutions as compared to the BH19 $N$-body simulation. As a consequence, the vorticity power spectrum is of similar magnitude as the velocity-divergence power spectrum at larger wave numbers and thus the dip indicating the scale of energy transfer is shifted towards higher wave numbers. An accurate reconstruction of the velocity-divergence power spectrum and the suppression in the mildly non-linear regime therefore necessitates a correct description of the vorticity power spectrum, which in the current framework requires a realistic velocity dispersion mean field at late times.

**Isotropic velocity dispersion equal-time auto-spectra.** In Figure 13, the time (upper panel) and wave number (lower panel) dependencies of the dimensionless isotropic velocity dispersion equal-time auto-spectrum are shown at constant wave number $k \approx 0.5\ h/\mathrm{Mpc}$ and at fixed redshift $z = 0$, respectively.

At constant wave number (upper panel) the power spectrum grows in time and does so more strongly for initially colder dark matter models. Also shown are three snapshots of the BH19 $N$-body simulation which exhibit a stronger growth at later times, roughly scaling $\propto D_+^8$. The numerical solutions cannot reproduce the time dependence observed in the BH19 $N$-body simulation since the time dependence is directly linked to the velocity dispersion mean field shown in Figure 10. The mean field fails to correctly reproduce the observed time dependence as initially rather warm dark matter models are studied here as discussed above.

At fixed redshift (lower panel) the power spectrum shows a power-law scaling in the infrared up to the free-streaming wave number. $N$-body simulations suggest that this scale is in fact related to shell-crossing and the largest collapsed structures [59, 70]. As indicated, the free-streaming wave number is related to the velocity dispersion mean field in the current framework. It is clearly visible that for wave numbers larger than the free-streaming scale the power spectrum is suppressed. The spectral index observed in $N$-body simulations is roughly $n_\sigma \approx -1$ [59] and corresponds to the indicated scaling $\Delta_{33} \propto k^6$.[19] The numerical solutions obtained here show a slightly less steep power spectrum that scales roughly as $\Delta_{33} \propto k^{5.2}$ in the infrared, before dropping at wave numbers above the free-streaming scale.

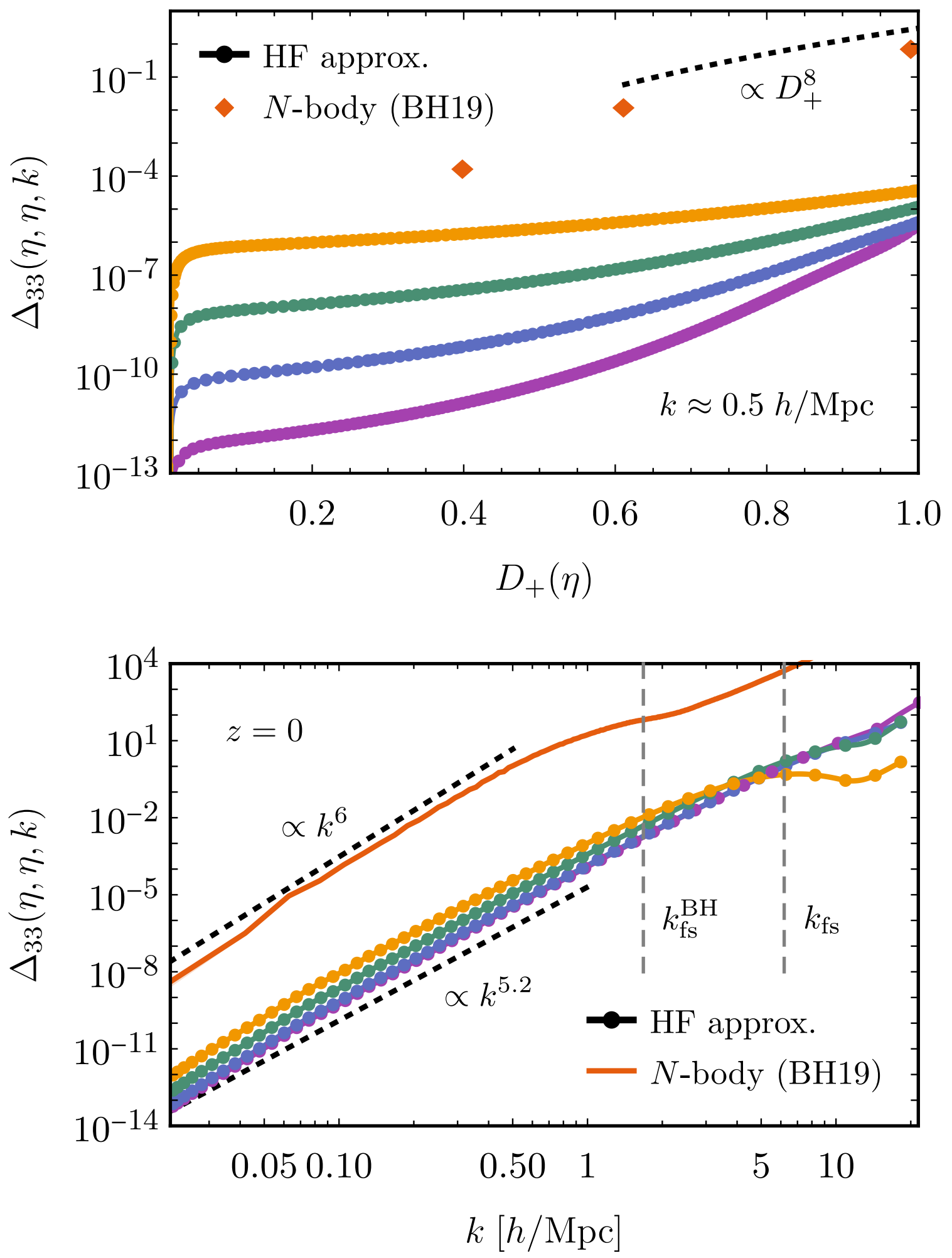

**Figure 13**. Time (upper panel) and wave number (lower panel) dependence of the dimensionless isotropic velocity dispersion equal-time auto-spectrum at constant wave number $k \approx 0.5\ h/\mathrm{Mpc}$ and redshift $z = 0$, respectively. The numerical solutions of the Hartree–Fock (HF) approximation are colour coded according to the respective mean fields shown in Figure 10. Three snapshots from data of the BH19 $N$-body simulation (red diamonds and red solid curves) as well as power-law scalings (black dotted lines) are also shown. The velocity dispersion of colder dark matter candidates grows faster in time but all candidates show a near to power law scaling in wave number at redshift $z = 0$ corresponding to a spectral index $n_\sigma \approx -1.8$ up to the free-streaming wave number.

**Velocity- and velocity dispersion-density contrast equal-time cross-spectra.** In Figure 14, the wave number dependence of the dimensionless velocity-divergence-density contrast (upper panel) and velocity dispersion-density contrast (lower panel) equal-time cross-spectra are shown at redshift $z = 0$.

[19] Remember that $\Delta_{33} \propto k^3 P_{33} \propto k^7 P_{\sigma\sigma}$ due to rescaling the field content (3.1).

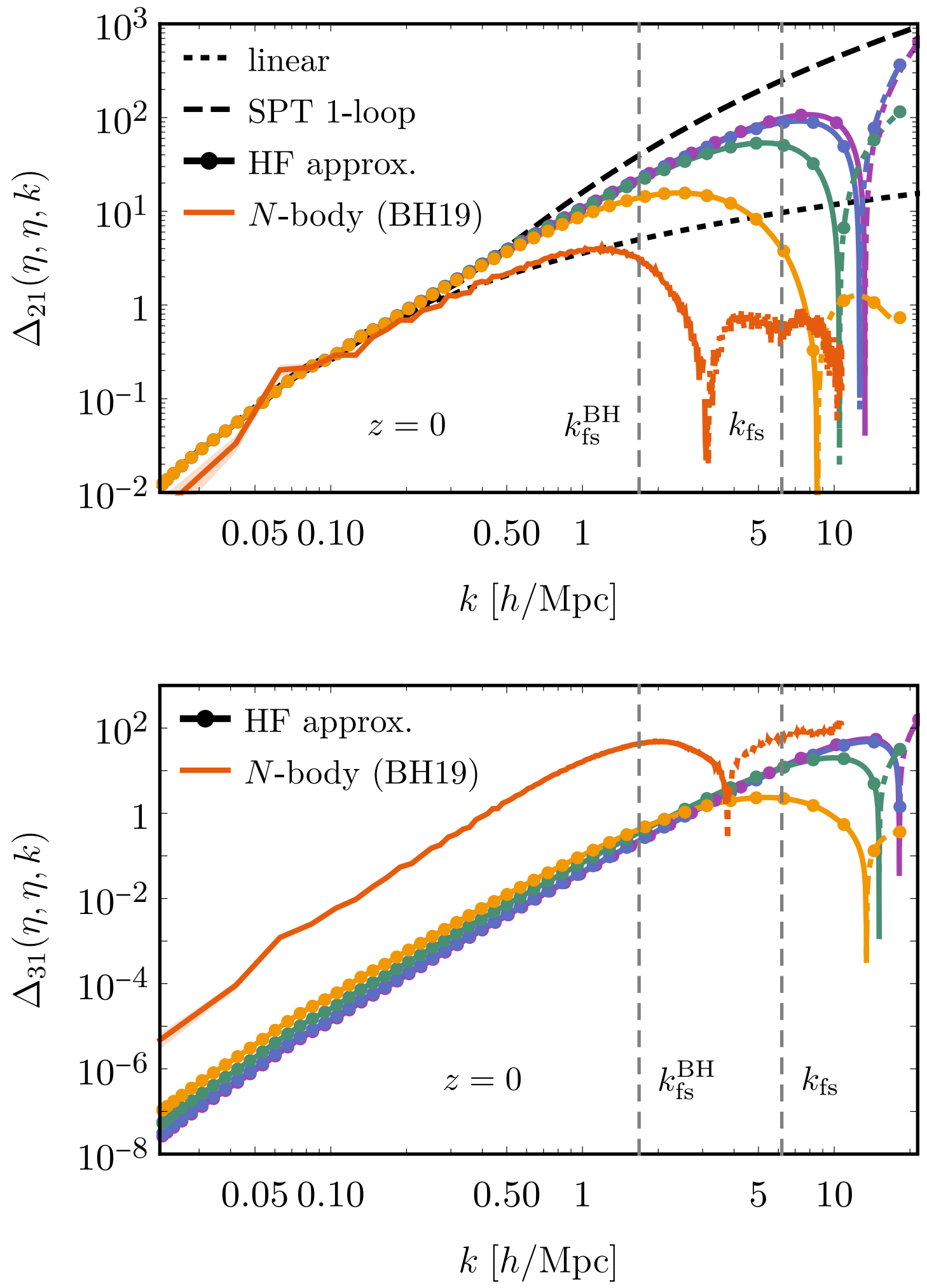

**Figure 14**. Dimensionless velocity-divergence- (upper panel) and velocity dispersion-density contrast (lower panel) equal-time cross-power spectrum $z = 0$. The numerical solutions of the Hartree–Fock (HF) approximation are colour coded according to the respective mean fields shown in Figure 10. They are compared to the linear power spectrum (black dotted curves), the standard perturbation theory (SPT) one-loop prediction (black dashed curves) and data from the BH19 $N$-body simulation (red solid curve). The sign change of the cross-spectra is associated with shell-crossing and captured by the Hartree–Fock approximation in both power spectra.

As seen in Figure 9, the single-stream approximation badly fails to describe the velocity-divergence-density contrast equal-time cross-spectrum. This is related to the fact that the sign change observed in $N$-body data is due to shell-crossing which cannot be captured by the single-stream approximation. In contrast, when including velocity dispersion degrees of freedom, the numerical solutions shown in Figure 14 exhibit the expected sign change. It is evident that the scale of this sign change is related to the free-streaming wave number (the sign change occurs roughly at $2k_{\rm fs}$) and thus to the velocity dispersion mean field. The numerical solutions shown here show a sign change at a higher wave number compared to the BH19 $N$-body simulation since the corresponding mean field is smaller in magnitude, as discussed above. Nonetheless, one can see that including velocity dispersion degrees of freedom into the description actually allows to describe effects associated with shell-crossing.

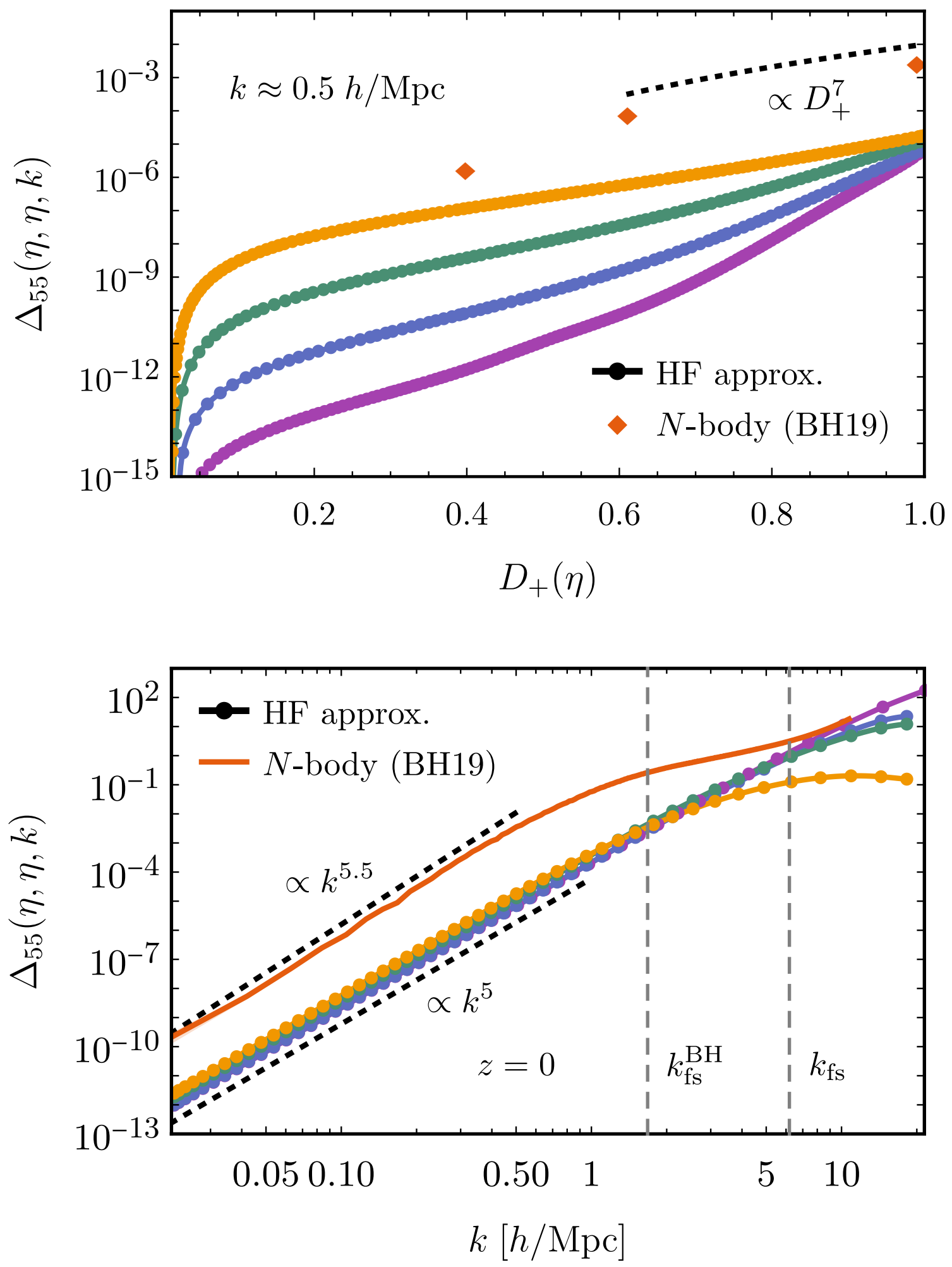

**Figure 15**. Time (upper panel) and wave number (lower panel) dependence of the dimensionless vorticity equal-time auto-spectrum at constant wave number $k \approx 0.5\ h$/Mpc and redshift $z = 0$, respectively. The numerical solutions of the Hartree–Fock (HF) approximation are colour coded according to the respective mean fields shown in Figure 10. Three snapshots from data of the BH19 $N$-body simulation (red diamonds and red solid curves) as well as power-law scalings (black dotted lines) are also shown. The power spectrum shows a near to power-law scaling in wave number at redshift $z = 0$ corresponding to a spectral index $n_\omega \approx 2$ up to the free-streaming wave number.

**Vorticity equal-time auto-spectra.** In Figure 15, the time (upper panel) and wave number (lower panel) dependencies of the dimensionless vorticity equal-time auto-spectrum are shown at constant wave number $k \approx 0.5\ h$/Mpc and fixed redshift $z = 0$, respectively.

In the numerical solutions shown here, vorticity is initially absent and can not be sourced by the mean field alone in the current framework, which is also the reason that it is absent in the Hartree approximations studied in Section 4. But it is sourced by the non-linear cou-pling of fluctuations and is therefore naturally generated in the Hartree–Fock approximation through the coupling of modes in the statistical self-energy.

One can observe the late-time power-law-like scaling in time $\propto D_+^7$ [14, 70] in the BH19 $N$-body data and the numerical solutions obtained here grow similarly, although depending on the specific dark matter model. This is related to the fact that vorticity is initially sourced by the coupling of density contrast and velocity dispersion fluctuations, the latter of which are sourced by the velocity dispersion mean field. Therefore, similar to the velocity dispersion power spectra shown in Figure 13, the time dependence of the vorticity power spectrum is sensitive to the evolution of the velocity dispersion mean field.

For the wave number dependence one finds a spectral index $n_\omega \approx 2.5$ in the infrared from various $N$-body simulations [14, 69, 70], corresponding to the scaling $\Delta_{55} \propto k^{5.5}$ of the dimensionless vorticity power spectrum, before dropping on scales smaller than the free-streaming scale. The numerical solutions obtained here show a spectral index of almost exactly $n_\omega \approx 2$.[20] Indeed, there are arguments that if vorticity is absent initially, one expects the power spectrum to actually scale with a spectral index $n_\omega = 2$. Since the position space covariance function only has finite support due to causality, the Fourier transform needs to be analytic [71]. Since the vorticity mode is transverse, the power spectrum carries an additional transverse projector $\mathrm{P}_{ij}(\boldsymbol{k})$ such that the simplest scaling required for analyticity is given by $n_\omega = 2$. Precisely such a scaling is observed in the infrared of the Hartree–Fock approximation. It has been argued [69] that a deviation from this scaling observed in $N$-body simulations could be due to missing contributions of large scale modes due to the finite simulation box size, rendering the $N$-body power spectrum too steep. Along similar lines one could argue that the deviation of the isotropic velocity dispersion equal-time auto-spectrum from $N$-body data shown in Figure 13 is due to this finite size effect.

It is remarked, that although the infrared is perturbative in the sense that fluctuations are small and the density contrast power spectrum is well described by perturbation theory, the generation of vorticity and velocity dispersion is highly non-perturbative. This is also the reason why perturbative methods do not correctly capture the scaling of the vorticity power spectrum [6, 72]. More recently, there have been investigations that reproduce a spectral index $n_\omega = 2$ at two-loop order in an approximation that involves also higher-order cumulants [9]. It is interesting that the vorticity power spectrum shown in the lower panel of Figure 15 is enhanced in the ultraviolet, while the velocity-divergence power spectrum shown in the lower panel of Figure 12 significantly drops below the linear prediction with a pronounced dip near the free-stream wave number. Indeed, when looking at the dimensionful power spectra $P_{\theta\theta}$ and $P_{\omega\omega}$, one actually finds that the peak of the vorticity power spectrum aligns with the position of the velocity-divergence power spectrum dip [70]. This is interpreted as a consequence of angular momentum conservation converting power from velocity-divergence to vorticity modes, preventing the further infall of matter [69].

## 6 Conclusions

In summary, the Dyson–Schwinger equations have been investigated in the Hartree and Hartree–Fock approximations for dynamical cosmological structure formation caused by dark matter. To this end, an extension of the perfect pressureless fluid dark matter model was studied by including velocity dispersion degrees of freedom and in particular a non-vanishing velocity dispersion mean field.

[20] Remember that the spectrum $P_{55}$ differs from the vorticity spectrum $P_{\omega\omega}$ only by wave number independent factors such that the spectral index is the same.

The inclusion of velocity dispersion modifies the equation of motion of the velocity field. In particular, a time-local 'Hartree-type' contribution to the self-energy causes the propagator to depend on the solution of the velocity dispersion mean field and introduces a wave number dependence associated to the free-streaming of dark matter particles. Further, a non-local 'Fock-type' contribution to the self-energy captures the non-linear coupling of modes through loop effects. The resulting coupled equations for the mean field, propagators and power spectra were solved numerically in the Hartree–Fock approximation. While this approximation has a self-consistent one-loop structure, loop corrections to the three- and higher-point 1PI vertices are neglected.

A crucial ingredient is the inclusion of a time-dependent velocity dispersion mean field. At early times the mean field is decaying due to the Hubble expansion, however, it typically reaches a turnover point at later times, after which it starts to grow as it becomes non-linearly sourced by fluctuations. The magnitude of this growth depends on the initial value, with smaller initial values resulting in larger growth, such that the final order of magnitude of the velocity dispersion mean field seems to be largely independent of the initial value. It is worth noting that the numerical solutions obtained here lead to values that are still smaller in magnitude than those observed in $N$-body simulations.

Overall, the numerical results agree well with expectations. More specifically, it was demonstrated that the equal-time density contrast power spectrum exhibits better convergence to $N$-body simulations compared to standard perturbation theory. Further, as long as dark matter is initially sufficiently cold, the strong growth of velocity dispersion at late times does not seem to significantly affect the equal-time density contrast power spectrum. This is drastically different for the velocity-divergence power spectrum. Although the equal-time velocity-divergence power spectrum could not be reconstructed accurately, the qualitative behaviour of the dip in the non-linear regime could be captured by including velocity dispersion degrees of freedom and is inherently tied to the failure of the single-stream approximation. At shell-crossing, vectorial velocity modes are generated and energy is transferred into these modes from scalar velocity modes. Therefore, the scale of the dip in the velocity-divergence power spectrum is related to the free-streaming wave number and in turn related to the shape of the vorticity power spectrum. The amplitude of the vorticity power spectrum is directly linked to the velocity dispersion mean field, making an accurate description of the mean field crucial for correctly reproducing the dip. However, the numerical solutions for the mean field obtained here are still smaller than those observed in $N$-body simulations. As a result, the vorticity spectrum has a smaller amplitude and consequently the dip in the velocity-divergence power spectrum appears at larger wave numbers compared to $N$-body simulations.

By including a non-vanishing velocity dispersion mean field, vorticity vector modes are naturally generated through the non-linear coupling of scalar modes, even if initially absent. The numerical solutions grow in time and exhibit a power-law wave number dependence of the equal-time power spectrum in the infrared with a spectral index $n_\omega \approx 2$, as expected from analytical arguments. This result is close to the spectra index $n_\omega \approx 2.5$ observed in $N$-Body simulations and agrees better than other (semi-)analytical methods with $N$-body data, although it should be stressed that more recently a spectral index $n_\omega = 2$ was reproduced in a perturbative higher-order cumulant approach [9].

Finally, the equal-time power spectrum for the trace of the velocity dispersion tensor was computed, starting from vanishing initial conditions. The velocity dispersion fluctuations are naturally sourced through the velocity dispersion mean field and the overall amplitude is therefore closely tied to the evolution of the mean field, very similar to the vorticity power spectrum. A power-law wave number dependence was found in the infrared with a spectral index of $n_\sigma \approx -1.8$, which is slightly below the spectral index $n_\sigma \approx -1$ observed in $N$-body simulations.

Due to computational limitation, only the scalar velocity dispersion degrees of freedom have been included in this work. Since the velocity dispersion vector and tensor modes can only be sourced non-linearly (if initially absent) through cross-correlations of velocity and velocity dispersion scalar modes, their impact is expected to be subdominant compared to the investigated scalar degrees of freedom. Further, the degree to which cold dark matter could be studied is also heavily limited by computational resources in the presented work. The dark matter candidates that were investigated are comparably warm as the colder candidates would necessitate an accurate resolution also far into the ultraviolet of the wave number regime. Here, the equations of motion are solved with a finite element method in wave numbers, which necessarily involves a rather large wave number grid for accurate results. For a precise treatment extending further into the ultraviolet, or even spanning the hole wave number domain, more advanced spectral methods might be more sensible to accurately capture the ultraviolet correlations sourcing the velocity dispersion mean field.

The findings of this work indicate that the velocity dispersion mean field is largely independent of the initial value, as it grows more strongly the colder the initial condition. The velocity and velocity dispersion power spectra are sensitive to the mean field, as it determines the free-streaming scale. Consequently, the velocity dispersion mean field is closely tied to the ultraviolet suppression of the power spectra, which in turn are responsible for the growth of the mean field. In other words, as the mean field becomes larger, the power spectra become more suppressed in the ultraviolet, resulting in a lesser contribution from fluctuations to the mean field. This chain of reasoning leads to the conjecture of a universal fixed point at late times for the velocity dispersion mean field.

It was shown, that for a viable description of dark matter, it is necessary to go beyond the single-stream approximation and perturbative methods. While the inclusion of velocity dispersion degrees of freedom allows for an effective multi-stream description that qualitatively captures effects related to shell-crossing, it is unclear whether the inclusion of higher-order velocity cumulants is needed for an accurate description of structure formation after shell-crossing. It is evident that non-perturbative methods are necessary to describe the non-linear regime, which can be understood from the perspective of an infinite but partial resummation of loop diagrams. Results of the self-consistent one-loop approximation suggest that corrections to higher-order 1PI vertices are needed to accurately describe even the mildly non-linear regime. To this end, the functional renormalisation group provides a method to systematically explore approximations and it is planned to investigate this direction further in the future.

## A Functional approach to cosmological field theory

Here, the Martin–Siggia–Rose/Janssen–de Dominicis response field formalism for cosmology is reviewed. For a detailed review of the functional formalism for the dynamics of stochastic systems, the reader is referred to reference [73].

Let $\psi_a(\eta, \boldsymbol{k})$ be a set of fields with statistically homogeneous and isotropic Gaussian random initial conditions, characterised by a mean field $\Psi_a^{\mathrm{in}}$ and power spectrum $P_{ab}^{\mathrm{in}}(k)$, and subject to the equations of motion

$$\left[\partial_\eta \delta_{ab} + \Omega_{ab}(\eta, k)\right] \psi_b(\eta, \boldsymbol{k}) + I_a(\eta, \boldsymbol{k}) = 0 \,, \tag{A.1}$$

where $\Omega_{ab}$ is field-independent and $I_a$ is an interaction term the is non-linear in the fields. With the Martin–Siggia–Rose formalism [74] one constructs the action

$$\begin{aligned} S[\psi, \hat{\psi}] = &- \mathrm{i} \int_{\eta, \boldsymbol{k}} \hat{\psi}_a(\eta, -\boldsymbol{k}) \left[ \left[\partial_\eta \delta_{ab} + \Omega_{ab}(\eta)\right] \psi_b(\eta, \boldsymbol{k}) + I_a(\eta, \boldsymbol{k}) \right] \\ &+ \int_{\boldsymbol{k}} \hat{\psi}_a(\eta_{\mathrm{in}}, -\boldsymbol{k}) \left[ \mathrm{i}\, \delta(\boldsymbol{k})\, \Psi_a^{\mathrm{in}} + \tfrac{1}{2} P_{ab}^{\mathrm{in}}(k)\, \hat{\psi}_b(\eta_{\mathrm{in}}, \boldsymbol{k}) \right] \,, \end{aligned} \tag{A.2}$$

by introducing a set of so-called response fields $\hat{\psi}_a(\eta, \boldsymbol{k})$.[21] Utilising the Janssen–de Dominicis formalism [75–77], one constructs the generating functional

$$Z[J, \hat{J}] = \int \mathcal{D}\psi \int \mathcal{D}\hat{\psi}\; \mathrm{e}^{-S + J_A \psi_A + \hat{J}_A \hat{\psi}_A} \,, \tag{A.3}$$

with source currents $J_a$ and $\hat{J}_a$. The measures $\mathcal{D}\psi$ and $\mathcal{D}\hat{\psi}$ are understood as the continuum limit of field configuration integrals on a lattice in time and space.[22] Uppercase letters from the beginning of the Latin alphabet denote DeWitt indices that run over the field content, time and space, i.e. $A = (a, \eta, \boldsymbol{k})$, and are summed and integrated over if appearing twice in a single term for discrete and continuous variables, respectively.

Correlation functions are obtained from the generating functional (A.3) by applying functional derivatives with respect to the source currents,

$$\langle \psi_{\boldsymbol{A}_1} \dots \psi_{\boldsymbol{A}_n} \rangle = \frac{1}{Z} \frac{\delta^n Z}{\delta J_{\boldsymbol{A}_1} \dots \delta J_{\boldsymbol{A}_n}} \,, \tag{A.4}$$

where boldface DeWitt indices additionally comprise the physical-response field structure, e.g. $J_{\boldsymbol{A}} = (J_A, \hat{J}_A)$. Physical correlations are obtained at vanishing source currents and are said to be 'on the equations of motion'.[23]

For the sake of brevity the notations

$$Z^{(n)}_{\boldsymbol{A}_1 \dots \boldsymbol{A}_n} = \frac{\delta^n Z}{\delta J_{\boldsymbol{A}_1} \dots \delta J_{\boldsymbol{A}_n}} \,, \qquad Z^{(m,n)}_{A_1 \dots B_n} = \frac{\delta^{m+n} Z}{\delta J_{A_1} \dots \delta J_{A_m} \delta \hat{J}_{B_1} \dots \delta \hat{J}_{B_n}} \,, \tag{A.5}$$

---

[21] Integrals over the entire time domain $\eta_{\mathrm{in}}$ to $\eta_{\mathrm{fi}}$ are abbreviated as

$$\int_\eta = \int_{\eta_{\mathrm{in}}}^{\eta_{\mathrm{fi}}} \mathrm{d}\eta \,.$$

[22] For the construction one assumes the existence of a unique solution of the bare equations of motion (A.1) in the time interval $\eta_{\mathrm{in}}$ to $\eta_{\mathrm{fi}}$. Given the stochastic nature of the initial conditions, the Itô prescription is employed for the discretization of stochastic differential equations.

[23] While general correlation functions are source-dependent, the expectation values introduced in Section 2 are on the equations of motion. For non-vanishing source currents the density contrast and velocity mean fields in equation (2.12) generally do not vanish.

are employed and similar for other generating functionals that are introduced in the following. The first- and second-order derivatives are related in this notation by

$$Z^{(1)}_{\boldsymbol{A}} = \begin{pmatrix} Z^{(1,0)}_{A} \\ Z^{(0,1)}_{A} \end{pmatrix} , \qquad Z^{(2)}_{\boldsymbol{AB}} = \begin{pmatrix} Z^{(2,0)}_{AB} & Z^{(1,1)}_{AB} \\ Z^{(1,1)}_{BA} & Z^{(0,2)}_{AB} \end{pmatrix} . \tag{A.6}$$

A general property of the Martin–Siggia–Rose/Janssen–de Dominicis formalism is the normalisation $Z[0, \hat{J}] = 1$, which implies that all correlation functions with respect to response fields alone vanish on the equations of motion.

In the following, it is convenient to work in terms of *connected correlation functions*, which are derived from the generating functional $W = \ln(Z)$ in the same manner as correlation functions are obtained from $Z$. The (source-depended) mean field is then obtained from $\langle \psi_{\boldsymbol{A}} \rangle = \Psi_{\boldsymbol{A}} = W^{(1)}_{\boldsymbol{A}}$.

While the generating functional (A.3) provides a formal way to obtain correlation functions, there is no obvious way to explicitly calculate them in a non-perturbative manner without solving the functional integral. In order to study non-perturbative approximation schemes, it is sensible to introduce yet another generating functional, namely the one-particle irreducible (1PI) effective action. It is defined as the Legendre transform of the generating functional of connected correlation functions with respect to the source currents,

$$\Gamma[\Psi] = \sup_{J,\hat{J}} \Big[ J_{\boldsymbol{A}} \Psi_{\boldsymbol{A}} - W[J] \Big] , \tag{A.7}$$

where the functional arguments are understood to run over the physical-response field structure. The 1PI effective action is the generating functional of *1PI correlation functions*, which are obtained by applying functional derivatives with respect to the mean fields $\Psi_{\boldsymbol{A}}$ in the same manner as correlation functions are obtained from the generating functional $Z$.[24]

Assuming the supremum on the right-hand side of equation (A.7) is obtained at some maximising (field-dependent) source currents, general properties of the Legendre transform imply that

$$\Gamma^{(1)}_{\boldsymbol{A}} = J_{\boldsymbol{A}} , \qquad \Gamma^{(2)}_{\boldsymbol{AB}} W^{(2)}_{\boldsymbol{BC}} = \delta_{\boldsymbol{AC}} . \tag{A.8}$$

When working in terms of connected and 1PI correlation functions, it is convenient to introduce a diagrammatic representation. To this end, denote

$W^{(m,n)}_{A_1 \dots A_m B_1 \dots B_n} =$ 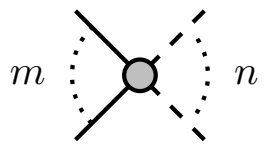 , $\Gamma^{(m,n)}_{A_1 \dots A_m B_1 \dots B_n} =$ 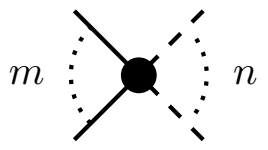 , (A.9)

where external solid (dashed) edges correspond to physical (response) source currents or fields, respectively. An edge of a connected correlation function can be joined to an edge of the same type (solid or dashed) of a 1PI correlation function, implying the summation and integration over the corresponding DeWitt indices. Due to the properties of $W$ and $\Gamma$, all expression of the form

$W^{(0,n)}_{B_1 \dots B_n} =$ 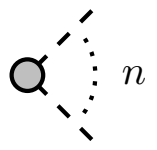 , $\Gamma^{(m,0)}_{A_1 \dots A_m} =$ 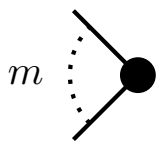 , (A.10)

vanish at vanishing source currents or on the equations of motion, respectively.

[24]Derivatives with respect to physical fields alone vanish on the equations of motion, $\Gamma^{(m,0)}_{A_1 \dots A_m} = 0$, since the 1PI effective action is at least linear in the response fields, which vanish on the equations of motion.

Diagrammatically, the equations for the 1PI one-point correlation functions read

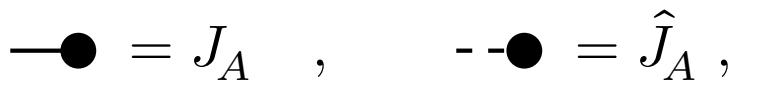

$$= J_A \quad , \qquad = \hat{J}_A \; , \tag{A.11}$$

while the equations for the 1PI two-point correlation functions are given by

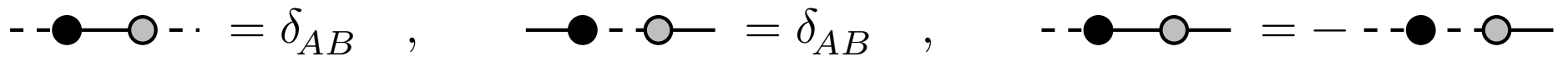

$$= \delta_{AB} \quad , \qquad = \delta_{AB} \quad , \qquad = - \quad . \tag{A.12}$$

To study approximation schemes, it is sensible to introduce a diagrammatic representation for the *linear connected* and *bare 1PI* correlation functions. These are represented by

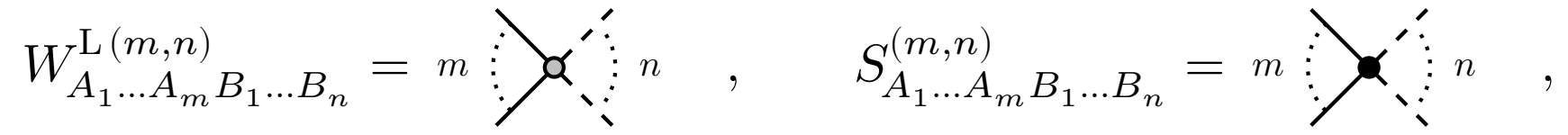

$$W^{\mathrm{L}\,(m,n)}_{A_1 \ldots A_m B_1 \ldots B_n} = \quad , \qquad S^{(m,n)}_{A_1 \ldots A_m B_1 \ldots B_n} = \quad , \tag{A.13}$$

similar to the full correlation functions but with smaller vertex symbols.

## B Bare vertices

The symmetrised vertices $\gamma_{abc}(\boldsymbol{k}_1, \boldsymbol{k}_2)$ are listed below, grouped by the outgoing mode with wave vector $\boldsymbol{k} = \boldsymbol{k}_1 + \boldsymbol{k}_2$.

Density contrast:

$$\begin{aligned}
\gamma_{121}(\boldsymbol{k}_1, \boldsymbol{k}_2) &= -\frac{\boldsymbol{k} \cdot \boldsymbol{k}_1}{2k_1^2} \; , \\
\gamma_{15_i1}(\boldsymbol{k}_1, \boldsymbol{k}_2) &= -\frac{(\boldsymbol{k}_1 \times \boldsymbol{k}_2)_i}{2k_1^2} \; .
\end{aligned} \tag{B.1}$$

Velocity-divergence:

$$\begin{aligned}
\gamma_{222}(\boldsymbol{k}_1, \boldsymbol{k}_2) &= -\frac{k^2\,(\boldsymbol{k}_1 \cdot \boldsymbol{k}_2)}{2k_1^2 k_2^2} \; , \\
\gamma_{225_i}(\boldsymbol{k}_1, \boldsymbol{k}_2) &= -\frac{[k_1^2 + 2(\boldsymbol{k}_1 \cdot \boldsymbol{k}_2)](\boldsymbol{k}_1 \times \boldsymbol{k}_2)_i}{2k_1^2 k_2^2} \; , \\
\gamma_{25_i5_j}(\boldsymbol{k}_1, \boldsymbol{k}_2) &= \frac{(\boldsymbol{k}_1 \times \boldsymbol{k}_2)_i (\boldsymbol{k}_1 \times \boldsymbol{k}_2)_j}{k_1^2 k_2^2} \; , \\
\gamma_{231}(\boldsymbol{k}_1, \boldsymbol{k}_2) &= \frac{\boldsymbol{k} \cdot \boldsymbol{k}_2}{2k_1^2} \; , \\
\gamma_{241}(\boldsymbol{k}_1, \boldsymbol{k}_2) &= \frac{3(\boldsymbol{k} \cdot \boldsymbol{k}_1)(\boldsymbol{k}_1 \cdot \boldsymbol{k}_2) - k_1^2(\boldsymbol{k} \cdot \boldsymbol{k}_2)}{4k_1^4} \; .
\end{aligned} \tag{B.2}$$

Isotropic velocity dispersion:

$$\begin{aligned}
\gamma_{332}(\boldsymbol{k}_1, \boldsymbol{k}_2) &= -\frac{3k^2(\boldsymbol{k}_1 \cdot \boldsymbol{k}_2) + 2k^2 k_2^2}{6k_1^2 k_2^2} \; , \\
\gamma_{335_i}(\boldsymbol{k}_1, \boldsymbol{k}_2) &= -\frac{k^2(\boldsymbol{k}_1 \times \boldsymbol{k}_2)_i}{2k_1^2 k_2^2} \; , \\
\gamma_{342}(\boldsymbol{k}_1, \boldsymbol{k}_2) &= \frac{k^2 k_1^2 k_2^2 - 3k^2(\boldsymbol{k}_1 \cdot \boldsymbol{k}_2)^2}{6k_1^4 k_2^2} \; , \\
\gamma_{345_i}(\boldsymbol{k}_1, \boldsymbol{k}_2) &= -\frac{k^2(\boldsymbol{k}_1 \cdot \boldsymbol{k}_2)(\boldsymbol{k}_1 \times \boldsymbol{k}_2)_i}{2k_1^4 k_2^2} \; .
\end{aligned} \tag{B.3}$$

Anisotropic velocity dispersion:

$$
\begin{aligned}
\gamma_{432}(\boldsymbol{k}_1,\boldsymbol{k}_2) &= \frac{k^2k_2^2-3(\boldsymbol{k}\cdot\boldsymbol{k}_2)^2}{3k_1^2k_2^2}\,, \\
\gamma_{435_i}(\boldsymbol{k}_1,\boldsymbol{k}_2) &= -\frac{(\boldsymbol{k}\cdot\boldsymbol{k}_2)(\boldsymbol{k}_1\times\boldsymbol{k}_2)_i}{k_1^2k_2^2}\,, \\
\gamma_{442}(\boldsymbol{k}_1,\boldsymbol{k}_2) &= -\frac{18(\boldsymbol{k}\cdot\boldsymbol{k}_1)(\boldsymbol{k}\cdot\boldsymbol{k}_2)(\boldsymbol{k}_1\cdot\boldsymbol{k}_2)+9(\boldsymbol{k}\cdot\boldsymbol{k}_1)^2(\boldsymbol{k}_1\cdot\boldsymbol{k}_2)}{12k_1^4k_2^2} \\
&\quad+\frac{6k^2(\boldsymbol{k}_1\cdot\boldsymbol{k}_2)^2+6k_1^2(\boldsymbol{k}\cdot\boldsymbol{k}_2)^2+3k^2k_1^2(\boldsymbol{k}_1\cdot\boldsymbol{k}_2)-2k^2k_1^2k_2^2}{12k_1^4k_2^2}\,, \\
\gamma_{445_i}(\boldsymbol{k}_1,\boldsymbol{k}_2) &= -\frac{[6(\boldsymbol{k}\cdot\boldsymbol{k}_1)(\boldsymbol{k}_1\cdot\boldsymbol{k}_2)+3(\boldsymbol{k}\cdot\boldsymbol{k}_1)^2-2k^2(\boldsymbol{k}_1\cdot\boldsymbol{k}_2)](\boldsymbol{k}_1\times\boldsymbol{k}_2)_i}{4k_1^4k_2^2} \\
&\quad+\frac{[2k_1^2(\boldsymbol{k}\cdot\boldsymbol{k}_2)+k^2k_1^2](\boldsymbol{k}_1\times\boldsymbol{k}_2)_i}{4k_1^4k_2^2}\,.
\end{aligned}
\tag{B.4}
$$

Vorticity mode:[25]

$$
\begin{aligned}
\gamma_{5_i25_j}(\boldsymbol{k}_1,\boldsymbol{k}_2) &= \frac{k_{1,i}k_{1,j}-\delta_{ij}(\boldsymbol{k}\cdot\boldsymbol{k}_1)}{2k_1^2}\,, \\
\gamma_{5_i5_j5_k}(\boldsymbol{k}_1,\boldsymbol{k}_2) &= \frac{[k_2^2-k_1^2](\boldsymbol{k}_1\times\boldsymbol{k}_2)_i\delta_{jk}-k_2^2\epsilon_{ijl}k_lk_{1,k}-k_1^2\epsilon_{ikl}k_lk_{2,j}}{2k_1^2k_2^2}\,, \\
\gamma_{5_i31}(\boldsymbol{k}_1,\boldsymbol{k}_2) &= -\frac{(\boldsymbol{k}_1\times\boldsymbol{k}_2)_i}{2k_1^2}\,, \\
\gamma_{5_i41}(\boldsymbol{k}_1,\boldsymbol{k}_2) &= \frac{[k_1^2+3(\boldsymbol{k}_1\cdot\boldsymbol{k}_2)](\boldsymbol{k}_1\times\boldsymbol{k}_2)_i}{4k_1^4}\,.
\end{aligned}
\tag{B.5}
$$

## Acknowledgments

This work is supported by the Deutsche Forschungsgemeinschaft (German Research Foundation) under Germany's Excellence Strategy and the Cluster of Excellence EXC 2181 (STRUCTURES), the Collaborative Research Centre SFB 1225 (ISOQUANT) as well as the research grant FL 736/3-1.

## References

[1] P. J. E. Peebles, *The Large-Scale Structure of the Universe*, Princeton University Press (1980).

[2] M. Davis & P. J. E. Peebles, *On the integration of the BBGKY equations for the development of strongly nonlinear clustering in an expanding universe*, Astrophys. J. Suppl. Ser. **34** (1977) 425.

[3] T. Buchert & A. Domínguez, *Modeling multi-stream flow in collisionless matter: approximations for large-scale structure beyond shell-crossing*, Astron. Astrophys. **335** (1998) 395.

[4] P. McDonald, *How to generate a significant effective temperature for cold dark matter, from first principles*, J. Cosmol. Astropart. Phys. **04** (2011) 032.

[25] The vorticity mode vertices are calculated by utilising the vector identity $\epsilon_{ikl}\partial_k(u_j\partial_ju_l) = -\epsilon_{ijk}\partial_j\epsilon_{klm}u_l\omega_m$.

[5] M. Pietroni, G. Mangano, N. Saviano & M. Viel, *Coarse-grained cosmological perturbation theory*, J. Cosmol. Astropart. Phys. **01** (2012) 019.

[6] A. Aviles, *Dark matter dispersion tensor in perturbation theory*, Phys. Rev. D **93** (2016) 063517.

[7] A. Erschfeld & S. Floerchinger, *Evolution of dark matter velocity dispersion*, J. Cosmol. Astropart. Phys. **06** (2019) 039.

[8] M. Garny, D. Laxhuber & R. Scoccimarro, *Perturbation theory with dispersion and higher cumulants: Framework and linear theory*, Phys. Rev. D **107** (2023) 063539.

[9] M. Garny, D. Laxhuber & R. Scoccimarro, *Perturbation theory with dispersion and higher cumulants: Nonlinear regime*, Phys. Rev. D **107** (2023) 063540.

[10] S. V. Tassev, *The Helmholtz Hierarchy: phase space statistics of cold dark matter*, J. Cosmol. Astropart. Phys. **10** (2011) 022.

[11] P. McDonald & Z. Vlah, *Large-scale structure perturbation theory without losing stream crossing*, Phys. Rev. D **97** (2018) 023508.

[12] L. M. Widrow & N. Kaiser, *Using the Schrödinger equation to simulate collisionless matter*, Astrophys. J. Lett. **416** (1993) L71.

[13] C. Uhlemann, *Finding closure: approximating Vlasov-Poisson using finitely generated cumulants*, J. Cosmol. Astropart. Phys. **10** (2018) 030.

[14] S. Pueblas & R. Scoccimarro, *Generation of vorticity and velocity dispersion by orbit crossing*, Phys. Rev. D **80** (2009) 043504.

[15] M. Crocce & R. Scoccimarro, *Renormalized cosmological perturbation theory*, Phys. Rev. D **73** (2006) 063519.

[16] M. Crocce & R. Scoccimarro, *Memory of initial conditions in gravitational clustering*, Phys. Rev. D **73** (2006) 063520.

[17] M. Crocce & R. Scoccimarro, *Nonlinear evolution of baryon acoustic oscillations*, Phys. Rev. D **77** (2008) 023533.

[18] M. Pietroni, *Flowing with time: a new approach to non-linear cosmological perturbations*, J. Cosmol. Astropart. Phys. **10** (2008) 036.

[19] A. Taruya & T. Hiramatsu, *A Closure Theory for Nonlinear Evolution of Cosmological Power Spectra*, Astrophys. J. **674** (2008) 617.

[20] T. Hiramatsu & A. Taruya, *Chasing the nonlinear evolution of matter power spectrum with a numerical resummation method: Solution of closure equations*, Phys. Rev. D **79** (2009) 103526.

[21] A. Taruya, T. Nishimichi, S. Saito & T. Hiramatsu, *Nonlinear evolution of baryon acoustic oscillations from improved perturbation theory in real and redshift spaces*, Phys. Rev. D **80** (2009) 123503.

[22] S. Anselmi, S. Matarrese & M. Pietroni, *Next-to-leading resummations in cosmological perturbation theory*, J. Cosmol. Astropart. Phys. **06** (2011) 015.

[23] S. Anselmi & M. Pietroni, *Nonlinear power spectrum from resummed perturbation theory: a leap beyond the BAO scale*, J. Cosmol. Astropart. Phys. **12** (2012) 013.

[24] F. Bernardeau, N. Van de Rijt & F. Vernizzi, *Resummed propagators in multicomponent cosmic fluids with the eikonal approximation*, Phys. Rev. D **85** (2012) 063509.

[25] F. Bernardeau, N. Van de Rijt & F. Vernizzi, *Power spectra in the eikonal approximation with adiabatic and nonadiabatic modes*, Phys. Rev. D **87** (2013) 043530.

[26] D. Baumann, A. Nicolis, L. Senatore & M. Zaldarriaga, *Cosmological non-linearities as an effective fluid*, J. Cosmol. Astropart. Phys. **07** (2012) 051.

[27] J. J. M. Carrasco, M. P. Hertzberg & L. Senatore, *The effective field theory of cosmological large scale structures*, J. High Energy Phys. **09** (2012) 082.

[28] R. A. Porto, L. Senatore & M. Zaldarriaga, *The Lagrangian-space Effective Field Theory of large scale structures*, J. Cosmol. Astropart. Phys. **05** (2014) 022.

[29] D. Blas, S. Floerchinger, M. Garny, N. Tetradis & U. A. Wiedemann, *Large scale structure from viscous dark matter*, J. Cosmol. Astropart. Phys. **11** (2015) 049.

[30] P. McDonald, *Dark matter clustering: A simple renormalization group approach*, Phys. Rev. D **75** (2007) 043514.

[31] S. Matarrese & M. Pietroni, *Resumming cosmic perturbations*, J. Cosmol. Astropart. Phys. **06** (2007) 026.

[32] S. Matarrese and M. Pietroni, *Baryonic Acoustic Oscillations via the Renormalization Group*, Mod. Phys. Lett. A **23** (2008), 25-32.

[33] S. Floerchinger, M. Garny, N. Tetradis & U. A. Wiedemann, *Renormalization-group flow of the effective action of cosmological large-scale structures*, J. Cosmol. Astropart. Phys. **01** (2017) 048.

[34] S. Floerchinger, M. Garny, A. Katsis, N. Tetradis and U. A. Wiedemann, *The dark matter bispectrum from effective viscosity and one-particle irreducible vertices*, J. Cosmol. Astropart. Phys. **09** (2019), 047.

[35] A. Erschfeld & S. Floerchinger, *Cosmological functional renormalization group, extended Galilean invariance, and approximate solutions to the flow equations*, Phys. Rev. D **105** (2022) 023506.

[36] F. Bernardeau, S. Colombi, E. Gaztañaga & R. Scoccimarro, *Large-scale structure of the Universe and cosmological perturbation theory*, Phys. Rep. **367** (2002) 1.

[37] J. Binney & S. Tremaine, *Galactic Dynamics*, Princeton University Press (1987).

[38] J. M. Bardeen, J. R. Bond, N. Kaiser & A. S. Szalay, *The Statistics of Peaks of Gaussian Random Fields*, Astrophys. J. **304** (1986) 15.

[39] P. Watts & P. Coles, *Statistical cosmology with quadratic density fields*, Mon. Not. R. Astron. Soc. **338** (2003) 806.

[40] D. Boyanovsky & J. Wu, *Small scale aspects of warm dark matter: Power spectra and acoustic oscillations*, Phys. Rev. D **83** (2011) 043524.

[41] R. Scoccimarro, *A New Angle on Gravitational Clustering*, Ann. N. Y. Acad. Sci. **927** (2001) 13.

[42] P. Valageas, *A new approach to gravitational clustering: A path-integral formalism and large-N expansions*, Astron. Astrophys. **421** (2004) 23.

[43] P. Valageas, *Large-N expansions applied to gravitational clustering*, Astron. Astrophys. **465** (2007) 725.

[44] V. Silveira & I. Waga, *Decaying $\Lambda$ cosmologies and power spectrum*, Phys. Rev. D **50** (1994) 4890.

[45] F. J. Dyson, *The S Matrix in Quantum Electrodynamics*, Phys. Rev. **75** (1949) 1736.

[46] J. Schwinger, *On the Green's functions of quantized fields. I*, Proc. Natl. Acad. Sci. U.S.A. **37** (1951) 452.

[47] J. Schwinger, *On the Green's functions of quantized fields. II*, Proc. Natl. Acad. Sci. U.S.A. **37** (1951) 455.

[48] P. Valageas, *Using the Zeldovich dynamics to test expansion schemes*, Astron. Astrophys. **476** (2007) 31.

[49] F. Bernardeau & P. Valageas, *Eulerian and Lagrangian propagators for the adhesion model (Burgers dynamics)*, Phys. Rev. D **81** (2010) 043516.

[50] R. H. Kraichnan, *Dynamics of Nonlinear Stochastic Systems*, J. Math. Phys. **2** (1961) 124.

[51] R. H. Kraichnan, *Decay of Isotropic Turbulence in the Direct-Interaction Approximation*, Phys. Fluids **7** (1964) 1030.

[52] C. Runge, *Ueber die numerische Auflösung von Differentialgleichungen*, Math. Ann. **46** (1895) 167.

[53] W. Kutta, *Beitrag zur näherungsweisen Integration totaler Differentialgleichungen*, Z. Math. Phys. **46** (1901) 435.

[54] J. R. Cash & A. H. Karp, *A Variable Order Runge-Kutta Method for Initial Value Problems with Rapidly Varying Right-Hand Sides*, ACM Trans. Math. Softw. **16** (1990) 201.

[55] C. Lubich, *Runge-Kutta Theory for Volterra Integrodifferential Equations*, Numer. Math. **40** (1982) 119.

[56] J. Kim, C. Park, G. Rossi, S. M. Lee & J. R. Gott III, *The New Horizon Run Cosmological N-Body Simulations*, J. Korean Astron. Soc. **44** (2011) 217.

[57] E. Komatsu et al., *Five-Year Wilkinson Microwave Anisotropy Probe* Observations: Cosmological Interpretation*, Astrophys. J. Suppl. Ser. **180** (2009) 330.

[58] J. Dubinski, J. Kim, C. Park & R. Humble, *GOTPM: a parallel hybrid particle-mesh treecode*, New Astron. **9** (2004) 111.

[59] M. Buehlmann & O. Hahn, *Large-scale velocity dispersion and the cosmic web*, Mon. Not. R. Astron. Soc. **487** (2019) 228.

[60] Planck Collaboration, *Planck 2015 results XIII. Cosmological parameters*, Astron. Astrophys. **594** (2016) A13.

[61] V. Springel, *The cosmological simulation code `Gadget-2`*, Mon. Not. R. Astron. Soc. **364** (2005) 1105.

[62] D. Blas, J. Lesgourgues & T. Tram, *The Cosmic Linear Anisotropy Solving System (CLASS). Part II: Approximation schemes*, J. Cosmol. Astropart. Phys. **07** (2011) 034.

[63] R. W. Hockney & J. W. Eastwood, *Computer Simulation Using Particles*, McGraw-Hill (1981).

[64] S. Shandarin, S. Habib & K. Heitmann, *Cosmic web, multistream flows, and tessellations*, Phys. Rev. D **85** (2012) 083005.

[65] T. Abel, O. Hahn & R. Kaehler, *Tracing the dark matter sheet in phase space*, Mon. Not. R. Astron. Soc. **427** (2012) 61.

[66] Y. P. Jing, *Correcting for the Alias Effect When Measuring the Power Spectrum Using a Fast Fourier Transform*, Astrophys. J. **620** (2005) 559.

[67] D. Jeong & E. Komatsu, *Perturbation Theory Reloaded. II. Nonlinear Bias, Baryon Acoustic Oscillations, and Millennium Simulation in Real Space*, Astrophys. J. **691** (2009) 569.

[68] P. Valageas, *Expansion schemes for gravitational clustering: computing two-point and three-point functions*, Astron. Astrophys. **484** (2008) 79.

[69] G. Jelic-Cizmek, F. Lepori, J. Adamek & R. Durrer, *The generation of vorticity in cosmological N-body simulations*, J. Cosmol. Astropart. Phys. **09** (2018) 006.

[70] O. Hahn, R. E. Angulo & T. Abel, *The properties of cosmic velocity fields*, Mon. Not. R. Astron. Soc. **454** (2015) 3920.

[71] R. Durrer & C. Caprini, *Primordial magnetic fields and causality*, J. Cosmol. Astropart. Phys. **11** (2003) 010.

[72] G. Cusin, V. Tansella & R. Durrer, *Vorticity generation in the Universe: A perturbative approach*, Phys. Rev. D **95** (2017) 063527.

[73] J. A. Hertz, Y. Roudi & P. Sollich, *Path integral methods for the dynamics of stochastic and disordered systems*, J. Phys. A **50** (2017) 033001.

[74] P. C. Martin, E. D. Siggia & H. A. Rose, *Statistical Dynamics of Classical Systems*, Phys. Rev. A **8** (1973) 423.

[75] C. de Dominicis, *Techniques de renormalisation de la théorie des champs et dynamique des phénomènes critiques*, J. Phys. Colloq. **37** (1976) C1-247.

[76] H.-K. Janssen, *On a Lagrangean for Classical Field Dynamics and Renormalization Group Calculations of Dynamical Critical Properties*, Z. Phys. B **23** (1976) 377.

[77] C. de Dominicis & L. Peliti, *Field-theory renormalization and critical dynamics above $T_c$: Helium, antiferromagnets, and liquid-gas systems*, Phys. Rev. B **18** (1978) 353.